\documentclass[aps,prb,amsmath,twocolumn,amssymb,floatfixng,showpacs,
superscriptaddress,footinbib]{revtex4-1}
\pdfoutput=1
\usepackage[dvips]{graphics}
\usepackage{hyperref}
\usepackage{bm}
\usepackage{epsfig}
\usepackage{enumerate}
\usepackage{color}
\usepackage{graphicx}
\definecolor{dred}{rgb}{0.6,0,0}
\hypersetup{
    pdfnewwindow=true,      
    colorlinks=true,       
    linkcolor=magenta,          
    citecolor=blue,        
    filecolor=blue,      
    urlcolor=blue        
}

\newcommand{\bea}{\begin{eqnarray}}
\newcommand{\eea}{\end{eqnarray}}
\newcommand{\beq}{\begin{equation}}  
\newcommand{\eeq}{\end{equation}}
\newcommand{\non}{\nonumber} 
 
\newcommand{\eg}{{\it e.g., }}
\newcommand{\etal}{{\it et al. }}

\newcommand\ie{{\it{i.e.,~}}}
\newcommand\tc{{TC~}}
\newcommand\sbk{{SkC~}}
\newcommand\fom{{FOM~}}
\newcommand\fomd{{FOM}}

\newcommand\up{\uparrow}
\newcommand\dn{\downarrow}

\newcommand\bdg{{\textsf{BdG~}}}
\begin{document}

\title{Thermoelectric properties of a ferromagnet-superconductor hybrid junction: Role of interfacial Rashba spin-orbit interaction}
\author{Paramita Dutta}
\email{paramitad@iopb.res.in}
\affiliation{Institute of Physics, Sachivalaya Marg, Bhubaneswar-751005, India} 
\author{Arijit Saha}
\email{arijit@iopb.res.in}
\affiliation{Institute of Physics, Sachivalaya Marg, Bhubaneswar-751005, India} 
\affiliation{Homi Bhabha National Institute, Training School Complex, Anushakti Nagar, Mumbai 400085, India}
\author{A. M. Jayannavar}
\email{jayan@iopb.res.in}
\affiliation{Institute of Physics, Sachivalaya Marg, Bhubaneswar-751005, India} 
\affiliation{Homi Bhabha National Institute, Training School Complex, Anushakti Nagar, Mumbai 400085, India}

\begin{abstract}
We investigate thermoelectric properties of a ferromagnet-superconductor hybrid structure with Rashba spin-orbit interaction and delta function potential barrier at the interfacial layer. The exponential rise of thermal conductance with temperature manifests a cross-over temperature scale separating two opposite behaviors of it with the change of polarization in the ferromagnet whereas the inclusion of interfacial Rashba spin-orbit field results in a non-monotonic behavior of it with the strength of Rashba field. We employ scattering matrix approach to determine the amplitudes of all the scattering processes possible at the interface to explain the thermoelectric properties of the device. We examine Seebeck effect and show that higher thermopower can be achieved when the polarization of the ferromagnet tends towards the half-metallic limit. It can be enhanced even for lower polarization in presence of the finite potential barrier. In presence of interfacial Rashba spin-orbit interaction, Seebeck coefficient rises with the increase of barrier strength and polarization at weak or moderate interfacial Rashba field. From the application perspective, we compute the figure of merit and show that $zT\sim 4-5$ with higher polarization of the ferromagnet both in absence and presence of weak or moderate Rashba spin-orbit interaction along with the scalar potential barrier.
\end{abstract}


\maketitle

\section{Introduction}
In comparison to the metallic and semiconducting material, the thermoelectric effects are strongly suppressed in 
superconductors~\cite{chandrasekhar2009,machon}. One of the reasons behind this is the interference of temperature 
dependent super current with the thermal current. The electron-hole symmetry present in the superconducting density 
of states (DOS) makes the opposite directional electron and hole thermo-currents (generated due to the thermal 
gradient) nullify each other~\cite{ozaeta2014}. 

Recently, superconducting hybrid structures, especially ferromagnet-superconductor (FS) junctions, have attracted a 
lot of research interests due to the dramatic boosts of thermoelectric effects in them~\cite{chandrasekhar2009,
kalenkov2012theory,ozaeta2014,machon,machon2,kolenda2,kolenda2016}. 
Inducing spin-triplet correlation within the superconductor and the asymmetric DOS profiles corresponding to the two spin 
sub-bands of the ferromagnet are the key features to be utilized in order to make such FS junction suitable in the context 
of thermal transport. The asymmetry in the two spin sub-bands according to the polarization of the ferromagnet can manipulate 
the Andreev reflection (AR) which occurs when an incoming electron reflects back as a hole from the FS interface resulting 
in a cooper pair transmission into the superconductor within the sub-gapped regime~\cite{andreev1}. Mixing of electron and 
hole-like excitations due to Andreev reflection may yield large electron-hole asymmetry. This asymmetry makes the expression 
of the thermoelectric coefficient to get rid of $E/T_F$ factor which is responsible for the low value of the thermoelectric 
coefficient in the normal state of the material~\cite{abrikosov1988fundamentals}.

In order to investigate the thermoelectric properties of a material or hybrid junction it is customary to derive the thermal 
conductance (TC) or thermal current generated by the temperature gradient~\cite{yokoyama2008heat,beiranvand2016spin,
alomar2014thermoelectric}. Particularly in case of superconducting hybrid junction, the information of the superconducting gap 
parameter like its magnitude, pairing symmetry etc. can be extracted from the behavior of the thermal conductance~\cite{yokoyama2008heat}. 
From the application perspective, it is more favorable to compute the Seebeck coefficient (SkC), known as 
thermopower, which is the open circuit voltage developed across the junction due to the electron flow caused by the thermal 
gradient~\cite{blundell2009concepts}. Enhancement of \sbk can pave the way of promising application to make an efficient 
heat-to-energy converter which may be a step forward to the fulfilment of the global demand of 
energy~\cite{alomar2014thermoelectric}. Since last few decades intense research is being carried out in search of newer and 
efficient energy harvesting devices that convert waste heat into electricity~\cite{hwang2015large,snyder2008complex}. Usage 
of good thermoelectric material is one of the ways of making those devices more efficient. Now in order to determine how 
good thermoelectric a system is, one can calculate \sbk as well as the dimensionless parameter called figure of merit (FOM) 
which is naively the ratio of the power extracted from the device to the power we have to continually provide in order to 
maintain the temperature difference~\cite{sevinccli2010enhanced,zebarjadi2012perspectives}. It provides us an estimate of 
the efficiency of a mesoscopic thermoelectric device like refrigerator, generator etc. based on thermoelectric 
effects~\cite{giazotto2006opportunities}. Improving this thermoelectric \fom with enhanced \sbk so that the heat-electricity
conversion is more efficient~\cite{goldsmid3electronic,xu2014enhanced,liu2010enhancement,dragoman2007giant,ohta2007giant}
is one of the greatest challenges in material science. Particularly, 
enhancement of the performance of any superconducting hybrid junction is much more challenging due to the above-mentioned reasons. 

The prospects of FS junction, as far as thermoelectric property is concerned, depend on the new ingredients to manipulate 
the spin dependent particle-hole symmetry. The latter has been implemented using external magnetic 
field~\cite{linder2016,ozaeta2014,bathen2016,kolenda2016,linder2007spin}, quantum dot at the junction~\cite{hwang}, non-uniform 
exchange field~\cite{alidoust2010phase}, phase modulation~\cite{zhao2003phase}, magnetic impurities~\cite{kalenkov2012theory} 
or internal properties like inverse proximity effect~\cite{peltonen2010} etc. Recently, Machon \etal~have considered simultaneous 
effects of spin splitting and spin polarized transport~\cite{machon} in order to obtain enhanced thermoelectric effects in FS 
hybrid structure. In addition to these effects, presence of spin-orbit field~\cite{linder2016,alomar2014thermoelectric} can 
play a vital role in this context. 

Study of interfacial spin-orbit coupling effect on transport phenomena has become a topic of intense research interest during 
past few decades due to the spin manipulation~\cite{datta1990electronic}. Interplay of the polarization and the interfacial 
field may lead to marked anisotropy in the junction electrical conductance~\cite{hogl} and Josephson current~\cite{costa16}. 
Interfacial spin-orbit field, especially Rashba spin-orbit field~\cite{rashba,rashba2} arising due to the confinement potential 
at the semiconductor or superconductor hybrid structure, can also be the key ingredient behind such spin 
manipulation~\cite{sun2015general}. 

The aspect of thermal transport in FS hybrid junction incorporating the role of interfacial spin-orbit interaction has not 
been studied in detail so far in case of ordinary ferromagnet. A few groups have performed their research in this direction 
in graphene~\cite{alomar2014thermoelectric,beiranvand2016spin}. Motivated by these facts, in this article we study 
thermoelectric properties of a FS structure with Rashba spin-orbit interaction (RSOI)~\cite{rashba,review} at the interfacial 
layer. We employ Blonder-Tinkham-Klapwijk~\cite{blonder} (BTK) formalism to compute the \tc, \sbk and \fom therein. We 
investigate the role of RSOI on the thermoelectric properties. The interfacial scalar barrier at the FS interface reduces 
\tc. On the other hand, the presence of RSOI at the FS interface can stimulate enhancement of \tc driven by the thermal 
gradient across the junction. In order to reveal the local thermoelectric response we investigate the behavior of the 
thermopower with the polarization, temperature as well as the barrier strength. \sbk is enhanced when the polarization of 
the ferromagnet increases towards the half-metallic limit. In presence of finite barrier at the junction, it could be higher 
even for low polarization. Presence of RSOI at the interface may reduce or enhance it depending on the barrier strength, 
temperature and the polarization. For higher barrier strength it always shows non-monotonic behavior with the temperature 
both in presence and absence of RSOI. Similar non-monotonic behavior is obtained for \fom with the rise of temperature and 
Rashba strength. We predict that FOM can exceed the value $1$ with higher polarization of the ferromagnet. The magnitude 
can even be more than $5$ for higher strength of barrier potential at the junction. It is also true in presence of weak 
RSOI. On the contrary, strong Rashba interaction can reduce it irrespective of the polarization and temperature. 

The remainder of the paper is organized as follows. In Sec.~\ref{modeltheory} we describe our model and the theoretical background. 
We discuss our results for thermal conductance, thermopower and Figure of merit in Sec.~\ref{result}. 
Finally, we summarise and conclude in Sec.~\ref{conclu}.

\section{Model and theoretical background}\label{modeltheory}
\begin{figure}[htb]
\begin{center}
\includegraphics[width=6.5cm,height=4.6cm]{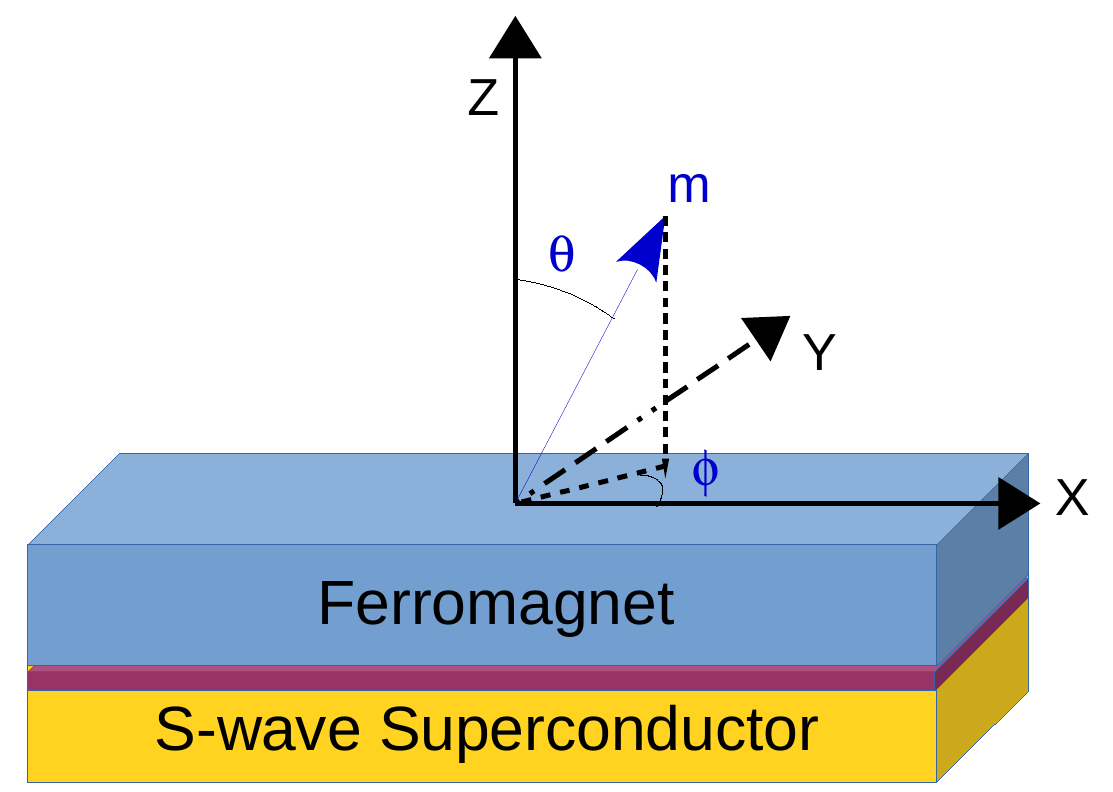}
\caption{(Color online) Cartoon of the FS junction with the magnetization vector $\mathbf{m}$. The dark red (dark grey) color is 
used to highlight the interfacial region of the FS hybrid structure. The F-region is kept at higher temperature ($T_F=T+\delta T/2$) 
compared to the S-region ($T_S=T-\delta T/2$) in order to maintain a temperature gradient ($\delta T=T_F-T_S$) across the junction.}
\label{geometry}
\end{center}
\end{figure}

We consider a model comprising of a ferromagnet F ($z>0$) and a $s$-wave superconductor S ($z<0$) hybrid structure as shown in 
Fig.~\ref{geometry}. The flat interface of semi-infinite ferromagnet-superconductor (FS) junction located at $z=0$ is modelled 
by a $\delta$-function potential with dimensionless barrier strength $Z$~\cite{vzutic2000tunneling,blonder} and Rashba spin-orbit 
interaction (RSOI) with strength $\lambda_{rso}$. The FS junction can be described by the Bogoliubov-deGennes (BdG) 
equation~\cite{de1999superconductivity} as,
\bea
\begin{bmatrix}
[\hat{H}_e-\mu]\hat{\sigma}_0 & \hat{\Delta}\\
\hat{\Delta}^\dagger & [\mu-\hat{H}_h]\hat{\sigma}_0 
\end{bmatrix} \Psi(\mathbf{r})=E \Psi(\mathbf{r})
\eea
where the single-particle Hamiltonian for the electron is given by,
\beq
\hat{H_e}=-(\hbar^2/2m)\nabla^2-(\Delta_{xc}/2) \Theta(z) \mathbf{m}.\hat{\mathbf{\sigma}}+\hat{H}_{int}.
\eeq
Similarly, for hole the Hamiltonian reads $\hat{H}_h=\hat{\sigma}_2 \hat{H}_e^*\hat{\sigma}_2$. The excitations of the electrons 
with effective mass $m$ are measured with respect to the chemical potential $\mu$. We set $m=1$ and $\mu=0$ throughout our 
calculation. The interfacial barrier is described by the Hamiltonian 
$\hat{H}_{int}=(V d\hat{\sigma_0}+ \mathbf{\omega}\cdot\hat{\sigma})\delta(z)$~\cite{hogl} with the 
height $V$, width $d$ and Rashba field $\omega$ $=\lambda[k_y,-k_x,0]$, $\lambda$ being the effective strength of the RSOI. 
The Stoner band model~\cite{stoner1939collective}, characterized by exchange spin splitting $\Delta_{xc}$, is employed to 
describe the F-region with the magnetization vector $\mathbf{m}=[\sin{\theta}\cos{\phi},\sin{\theta}\sin{\phi},\cos{\theta}]$. 
Here $\hat{\sigma}$ is the Pauli spin matrix. Note that, the growth direction ($z$-axis) of the heterostructure is chosen 
along [001] crystallographic axis~\cite{matos2009angular}. The superconducting pairing potential is expressed as 
$\hat{\Delta}=\Delta_s \Theta(z)\hat{\sigma_0}$. We assume it to be a spatially independent positive constant following 
Ref.~\onlinecite{hogl}.

\begin{figure}[htb]
\begin{center}
\includegraphics[width=7.7cm,height=4.5cm]{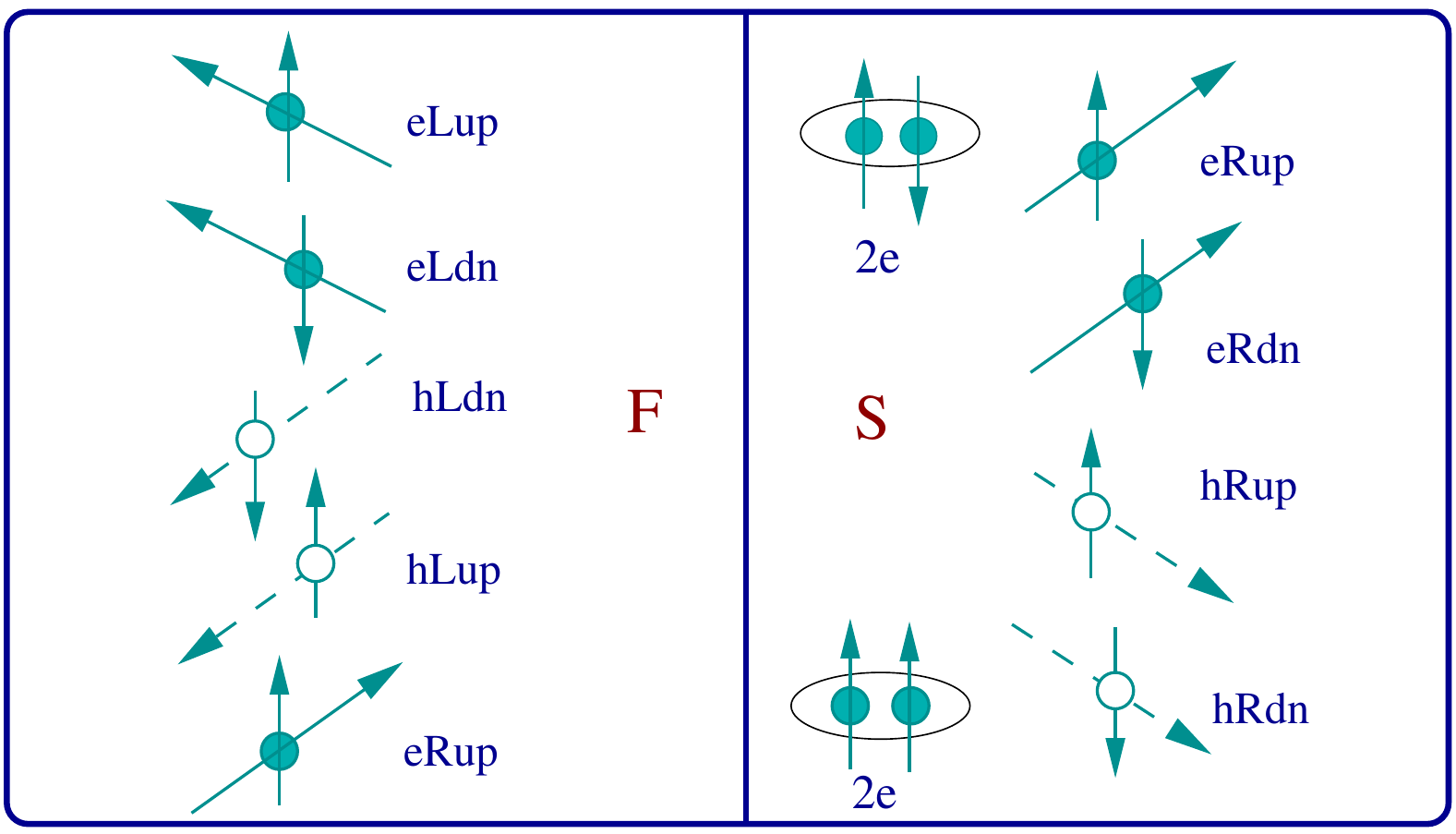}
\caption{(Color online) Schematic diagram for the quantum mechanical scattering processes taking place at FS interface. The solid and 
hollow spheres are used to denote electron (e) and hole (h), respectively. The letters `R (L)' indicates the right (left)-moving 
particles. Corresponding spin states are denoted by `up' ($\up$) and `down' ($\dn$), respectively.}
\label{scattering}
\end{center}
\end{figure}

Depending on the incoming electron energy there are four scattering processes possible at the FS interface. For electron 
with a particular spin, say $\sigma$, there can be normal reflection (NR), Andreev reflection (AR), tunneling as electron 
like (TE) or hole like (TH) quasi-particles. In addition to these phenomena there may be spin-flip scattering processes 
due to the interfacial spin-orbit field. Accordingly, we can have spin-flip counter parts of the above-mentioned four 
scattering processes namely, spin-flip NR (SNR), spin-flip AR (SAR), spin-flip TE (STE) and spin-flip TH 
(STH)~\cite{de1995andreev,cao2004spin}. The above mentioned scattering processes are schematically displayed in 
Fig.~\ref{scattering} 
for a right-moving electron with spin $\up$ (eRup). Note that, due to the possibility of spin-flip scattering processes in presence of
RSOI at the FS interface, spin-triplet~\cite{eschrig2011spin} superconducting correlation ($\up\up$ or $\dn\dn$) can be induced 
in addition to the conventional singlet pairing ($\up\dn$ or $\dn\up$)~\cite{hogl}.

The solution of the \bdg equations for the F-region, describing electrons and holes with spin $\sigma$, can be written as~\cite{hogl},
\bea
\Psi_{\sigma}^F(z)&=&\frac{1}{\sqrt{k_{\sigma}^e}} e^{i k_{\sigma}^e z}\psi_{\sigma}^e + r_{\sigma,\sigma}^e
e^{-ik^e_{\sigma}z}\psi_{\sigma}^e+r_{\sigma,\sigma}^h e^{ik^e_{\sigma}z}\psi_{\sigma}^h \non \\
&&+r_{\sigma,-\sigma}^e e^{-ik^e_{-\sigma}z}\psi_{-\sigma}^e 
+r_{\sigma,-\sigma}^e e^{ik^e_{-\sigma}z}\psi_{-\sigma}^e
\label{enormal}
\eea
where {\small $k^{e(h)}_{\sigma}=\sqrt{k_F^2-k_{||}^2+2m(\sigma\Delta_{xc}/2+(-) E)/\hbar^2}$} is the electron (hole)-like wave vector. 
$\sigma$ may be $\pm 1$ depending on whether the spin is parallel or anti-parallel to the vector $\mathbf{m}$. $k_F$ and $k_{||}$ are 
the Fermi and in-plane wave vector, respectively. The spinors for the electron-like and hole-like quasi-particles are respectively 
$\psi_{\sigma}^e=[\psi_{\sigma},0]^T$ and $\psi_{\sigma}^h=[0,\psi_{\sigma}]^T$ with 
{\small $\psi_{\sigma}^T=[\sigma\sqrt{1+\sigma\cos{\theta}}e^{-i \phi},\sqrt{1-\sigma\cos{\theta}}]/\sqrt{2}$}.
Here, $r^{e(h)}_{\sigma,\sigma^{\prime}}$ corresponds to the amplitude of normal (Andreev) reflection from the FS interface. $\sigma$ and 
$\sigma^{\prime}$ are the spin states for the incident and reflected electron or hole depending on the spin-conserving or spin-flipping 
process. Similarly, inside the superconducting region the solutions for the electron-like and hole like quasiparticles read~\cite{hogl} 
\bea
\Psi_{\sigma}^{S}=t^e_{\sigma,\sigma}\left[\begin{array}{c} u
\\ 0 \\v\\0\end{array} \right]{e}^{iq_{e}z}+ t^e_{\sigma,-\sigma}\left[\begin{array}{c} 0
\\ u \\0\\v\end{array} \right]{e}^{iq_{e}z}\non \\
+t^h_{\sigma,\sigma}\left[\begin{array}{c} u
\\ 0 \\v\\0\end{array} \right]{e}^{-iq_{h}z}+ t^h_{\sigma,-\sigma}\left[\begin{array}{c} 0
\\ u \\0\\v\end{array} \right]{e}^{-iq_{h}z},
\label{esupcon}
\eea
where the $z$-components of the quasi-particle wave vectors can be expressed as, 
{\small $q_{e(h)}=\sqrt{q_F^2-k_{||}^2+(-)2 m \sqrt{E^2-\Delta^2}/\hbar^2}$} and the superconducting coherence factors are 
{\small $u(v)=\sqrt{[1\pm\sqrt{1-\Delta^2/E^{2}}]/2}$}. We set the Fermi wave vector in both the F and S-regions to be the same 
\ie $q_F=k_F$~\cite{hogl}. Note that, we have written only the $z$ component of the wave functions. In the $x-y$ plane the wave 
vector is conserved giving rise to the planar wave function as,  $\Psi_{\sigma}(\mathbf{r})=\Psi_{\sigma}(z) e^{i (k_{x} x+k_{y} y)}$ 
where $k_x$ and $k_y$ are the components of $k_{||}$. Here, $t_{\sigma,\sigma^{\prime}}^{e(h)}$ denotes the amplitude of spin-conserving 
or spin-flipping transmitted electron (hole) like quasi-particles in the S region. We obtain the reflection and transmission amplitudes 
using the boundary conditions as,
\bea
\Psi^F_{\sigma}|_{z=0^+}&=&\Psi_{\sigma}^S|_{z=0^-}~,\non \\
\frac{\hbar^2}{2m}\left(\frac{d}{dz}\Psi_{\sigma}^S|_{z=0^-}-\frac{d}{dz}\zeta\Psi_{\sigma}^F|_{z=0^+}\right)
&=&V d~\zeta\Psi_{\sigma}^F|_{z=0^+} \non \\
+\begin{bmatrix} \mathbf{\omega}.\hat{\sigma} & 0 \\
                  0 & -\mathbf{\omega}.\hat{\sigma}
\end{bmatrix}
\Psi_{\sigma}^F|_{z=0^+}
\eea
where $\zeta=diag(1,1,-1,-1)$. We describe our results in terms of the dimensionless barrier strength $Z=\frac{V \ d \ m}{\hbar^2 k_F}$, 
RSOI strength $\lambda_{rso}=\frac{2m\lambda}{\hbar^2}$ and spin polarization $P=\frac{\Delta_{xc}}{2 E_F}$.\\

In presence of thermal gradient across the junction with no applied bias voltage, the electronic contribution to the thermal conductance,
(see appendix~\ref{appndx1} for details), in terms of the scattering processes is given by~\cite{blonder,yokoyama2008heat},
\bea
\kappa&=&\sum\limits_{\sigma}\int\limits_0^{\infty}\int\limits_{s}\frac{d^2k_{||}}{2 \pi k_F^2}\left[1-R^h_{\sigma}-R^e_{\sigma} \right] \non \\
&&~~~~~~~~~~~~~\left[\frac{(E-E_F)^2}{T^2\cosh^2{(\frac{E-E_F}{2k_B T})}}\right]dE 
\label{kappa_form}
\eea
where the NR and AR probability can be defined as 
$ R_{\sigma}^{e(h)}(E,k_{||})=Re[k_{\sigma}^{e(h)}|r_{\sigma}^{e(h)}|^2+k_{-\sigma}^{e(h)}|r_{-\sigma}^{e(h)}|^2]$ satisfying the current conservation. 
Here, the integration with respect to $k_{||}$ is performed over the entire plane $x-y$ of the interface. It is convenient to define a dimensionless 
wave vector $k=k_{||}/k_F$ and compute the integration in terms of it while calculating the TC. The Boltzmann constant is
denoted by $k_B$. $T$ is scaled by $T_c$, 
which is the critical temperature of the conventional singlet superconductor. 

Within the linear response regime, we obtain the expression for the thermopower or \sbk in unit of $k_B/e$ as follows~\cite{wysokinski2012thermoelectric},
\beq
S =-\left(\frac{V}{\delta T}\right)_{I=0}=-\frac{1}{T} \frac{\alpha}{G}
\label{sbk_exp}
\eeq
where the thermoelectric coefficient $\alpha$ and the electrical conductance $G$, in unit of $G_0$ ($e^2/h$), are represented as,
\bea
\alpha&=&\sum\limits_{\sigma}\int\limits_0^{\infty}\int\limits_{s}\frac{d^2k_{||}}{2 \pi k_F^2}\left[1-R^h_{\sigma}-R^e_{\sigma} \right] \non \\
&&~~~~~~~~~~~~~~~~~~\left[\frac{(E-E_F)}{T \cosh^2(\frac{E-E_F}{2 T})}\right]dE  
\label{alphaform}
\eea
and 
\bea
G&=&\sum\limits_{\sigma}\int\limits_0^{\infty}\int\limits_{s}\frac{d^2k_{||}}{2 \pi k_F^2}\frac{\left[1+R^h_{\sigma}-R^e_{\sigma}\right]}
{\left[T\cosh^2{(\frac{E-E_F}{2 T})}\right]}dE .
\label{intG}
\eea
Here $\alpha$ is expressed in unit of $G_0 k_B T/e$ ($\equiv k_B e T /h$). In terms of SkC, electrical conductance and thermal 
conductance, the \fom $zT$ is given by,
\bea
zT=\frac{S^2 G T}{K}
\label{merit}
\eea
where $K=\kappa-\frac{\alpha^2}{TG}$ is expressed in unit of $\kappa_0$ ($\equiv k_B^2T/h$). After applying the temperature difference 
between the two sides of the junction we obtain thermal current which essentially develops a voltage difference between them following 
the Peltier effect. This causes a correction to the thermal conductance as well. We consider such correction while defining the \fom of 
the system as every material manifesting Seebeck effect must exhibit the Peltier effect~\cite{bardas1995peltier}.

\section{Results and Discussion}\label{result}
In this section we present our numerical results for \tc, \sbk and \fom of the ferromagnet-superconductor junction, both in absence 
and presence of interfacial RSOI, in three different sub-sections. We discuss our results in terms of the scattering processes 
that occur at the interface of the FS hybrid structure and various parameters of the system. 
\subsection{Thermal conductance}
In this subsection we discuss the effect of polarization and RSOI, both in absence and presence of finite scalar barrier, on the behavior of 
the \tc throughout the temperature regime from low to high.
\subsubsection{Effect of polarization and barrier in absence of RSOI}
\begin{figure}[htb]
\begin{center}
\includegraphics[width=8.7cm,height=7.5cm]{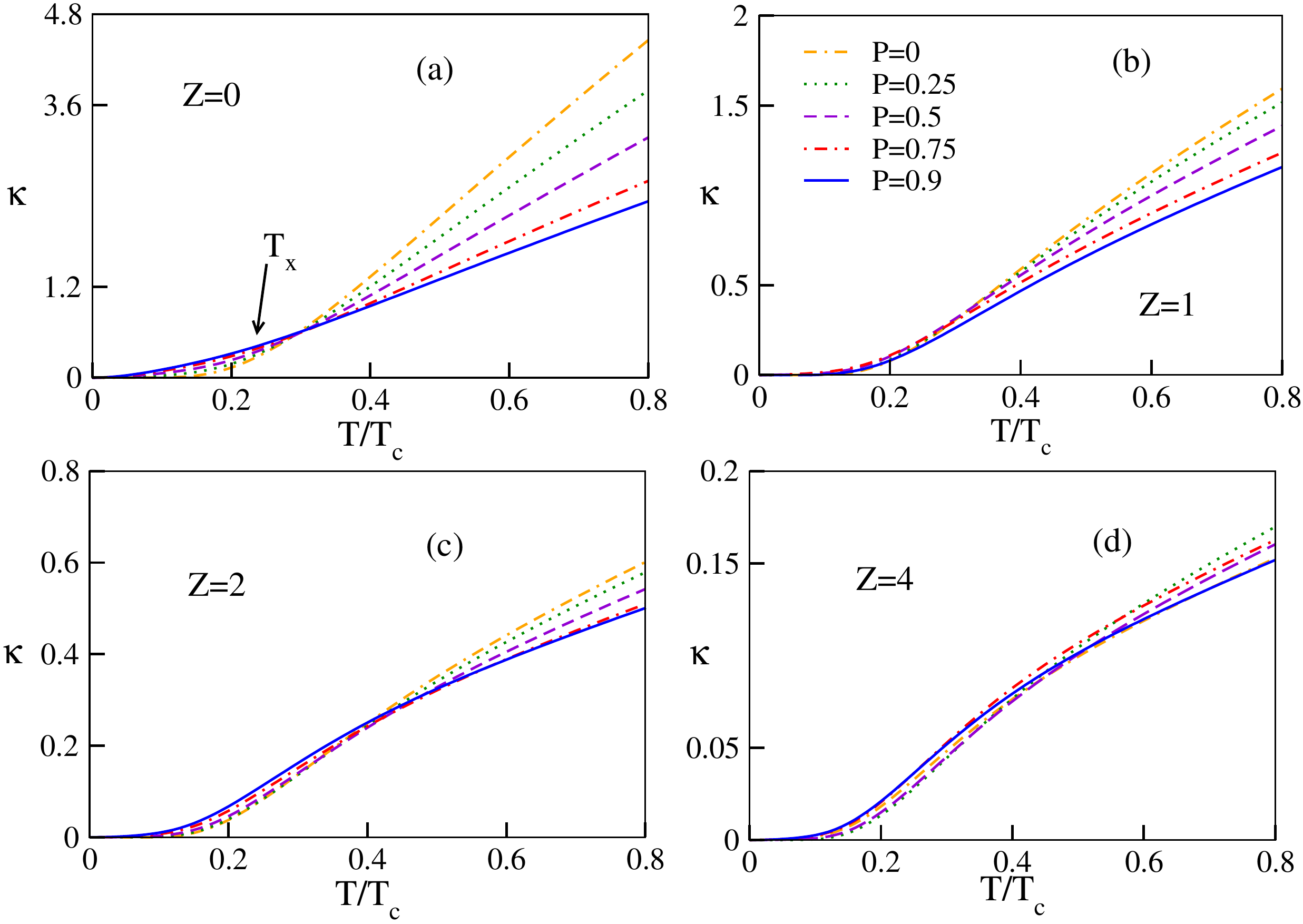}
\caption{(Color online) The behavior of thermal conductance ($\kappa$), in unit of $k_B^2/h$, is shown as a function of temperature 
($T/T_c$) in absence of RSOI ($\lambda_{rso}=0$) for different values of barrier strength ($Z$) and polarization ($P$) of the ferromagnet.}
\label{cond_T}
\end{center}
\end{figure}

In Fig.~\ref{cond_T} we show the variation of \tc $\kappa$ as a function of temperature $T/T_c$ in absence of RSOI 
for various polarization strength $P$ of the ferromagnet, starting from the unpolarized ($P=0$) towards the half-metallic ($P=0.9$) 
limit. Fig.~\ref{cond_T}[(a), (b), (c) and (d)] correspond to the interfacial scalar barrier strength $Z=0$, $1$, $2$ and $4$, respectively. 
From all the four figures it is apparent that \tc increases exponentially with temperature. This behavior, being independent of the barrier 
strength, is true for conventional normal metal-superconductor junction ($P=0$) as well as for any finite value of polarization ($P\ne 0$) 
of the ferromagnet. The fully developed gap of the superconductor is responsible for the exponential increase of the thermal 
conductance~\cite{andreev,andreev1} 
With the increase of temperature, the superconducting gap decreases resulting in reduction of AR amplitude and simultaneous increase of 
tunneling as electron-like quasi-particles. Thermal resistance of the superconductor falls off exponentially as the temperature is 
increased~\cite{andreev}. As a consequence, $\kappa$ rises with the temperature following an exponential nature. 

However, the rate of increase of $\kappa$ completely depends on the polarization $P$ of the ferromagnet. Gradual tunability of the 
polarization $P$ does not ensure any monotonic 
behavior of the TC. It depends on both the temperature and barrier strength. To illustrate this, we discuss the scenarios for 
different values of $Z$ one by one. When $Z=0$, the rate of increase of $\kappa$ is very slow with the increase of polarization for 
a particular value of $T/T_c$ (see Fig.~\ref{cond_T}(a)). This is true as long as $T/T_c<0.3$. On the other hand, for $T/T_c>0.3$ the scenario 
becomes opposite \ie~$\kappa$ starts decreasing with the change of polarization for a fixed $T/T_c$. There is a cross-over temperature 
$T_x$ ($\sim 0.3$ in this case) separating the two different behaviors of the \tc with polarization. We explain this phenomenon as 
follows. For very low $T/T_c$ \ie~$T<T_x$, superconductor gap parameter does not change by appreciable amount. In this situation, 
increase of polarization causes reduction of AR due to minority spin sub-band. This results in enhancement of \tc. Such enhancement is 
maximum towards the half-metallic limit ($P=0.9$) when AR vanishes due to the absence of minority spin-band. After a certain cross-over 
temperature, the gap decreases significantly with $T/T_c$ and the tunneling increases accordingly as long as $T \simeq T_c$. On top 
of that, if we increase the polarization, tunneling due to the minor spin band decreases leaving the major spin band contribution 
unchanged. As a whole, the change of behaviors of all the scattering processes results in reduction of \tc with polarization in the 
high temperature regime ($T>T_x$). 

Now let us consider finite $Z$ at the interface. In presence of the barrier, incident electrons encounter NR along with AR from the 
interface. NR reduces $\kappa$. Hence, the higher is the barrier strength $Z$, the lower is $\kappa$ for a particular temperature and 
polarization. This is apparent by comparing all the four figures of Fig.~\ref{cond_T}. The cross-over temperature $T_x$, separating the 
behaviors of the \tc with $P$, decreases as soon as we consider finite $Z$ as depicted in Fig.~\ref{cond_T}(b). It becomes $\sim$ 0.2 for 
$Z=1$. However, $T_x$ translates towards the high temperature limit with the increase of barrier strength (see Fig.~\ref{cond_T}(c) and (d)). 
For low $Z$, \tc does not change by appreciable amount with the increase of $P$ because of NR. In the low temperature regime, enhancement 
of $P$ causes reduction of AR. This does not ensure the increase of \tc as tunneling decreases due to the reflection from the interface. 
As $Z$ is enhanced, NR starts dominating over the other processes. This not only causes reduction of \tc but also translates $T_x$ towards 
the high temperature regime. For example, $T_x \sim 0.5$ (see Fig.~\ref{cond_T}(c)) and $0.8$ (see Fig.~\ref{cond_T}(d)) for $Z=2$ and 
$Z=4$, respectively. More over, there is a tendency of saturation of $\kappa$ when $T\rightarrow T_c$ irrespective of $P$ for higher barrier 
strength associated with very small change of $\kappa$ with $P$. For higher $Z$, AR, TE and TH are dominated by NR. Therefore, tuning 
polarization does not cause appreciable variation in the tunneling as well as AR resulting in very small change of \tc leading towards 
its saturation.

Therefore, the effect of polarization of the ferromagnet cannot be uniquely determined. The behavior of \tc with 
polarization changes depending on the temperature and the barrier strength as well.

So far, we have not discussed about the orientation of the magnetization. We present all of our results for $\theta=\pi/2$ and $\phi=0$. 
Very recently, H\"{o}gl \etal~have revealed the fact that electronic conductance shows an anisotropy with the rotation of the magnetization 
$\mathbf{m}$~\cite{hogl}. However, in case of thermal transport, contributions from all the energy values are taken into consideration. 
Therefore, with the change of $\mathbf{m}$ there is no appreciable change in \tc as all contributions due to different 
orientations of $\mathbf{m}$ are averaged out during integration over the energy (see Eq.(\ref{kappa_form})). This fact remains unchanged 
for any temperature ($T<T_c$) and polarization $P$.
\subsubsection{Interplay of polarization, barrier and interfacial RSOI}
Here we incorporate RSOI at the interface of the FS junction. The behavior of the \tc with the polarization and temperature in presence 
of RSOI remain qualitatively similar to that in absence of RSOI.  We refer to Appendix~\ref{appndx2} for more details manifesting this qualitative 
similarity. On the other hand, the magnitude of $\kappa$ may increase or decrease in presence of RSOI. It completely depends on the 
strength of RSOI as well as the strength of the barrier at the interface. 

\begin{figure}[htb]
\begin{center}
\includegraphics[width=8.7cm,height=7.3cm]{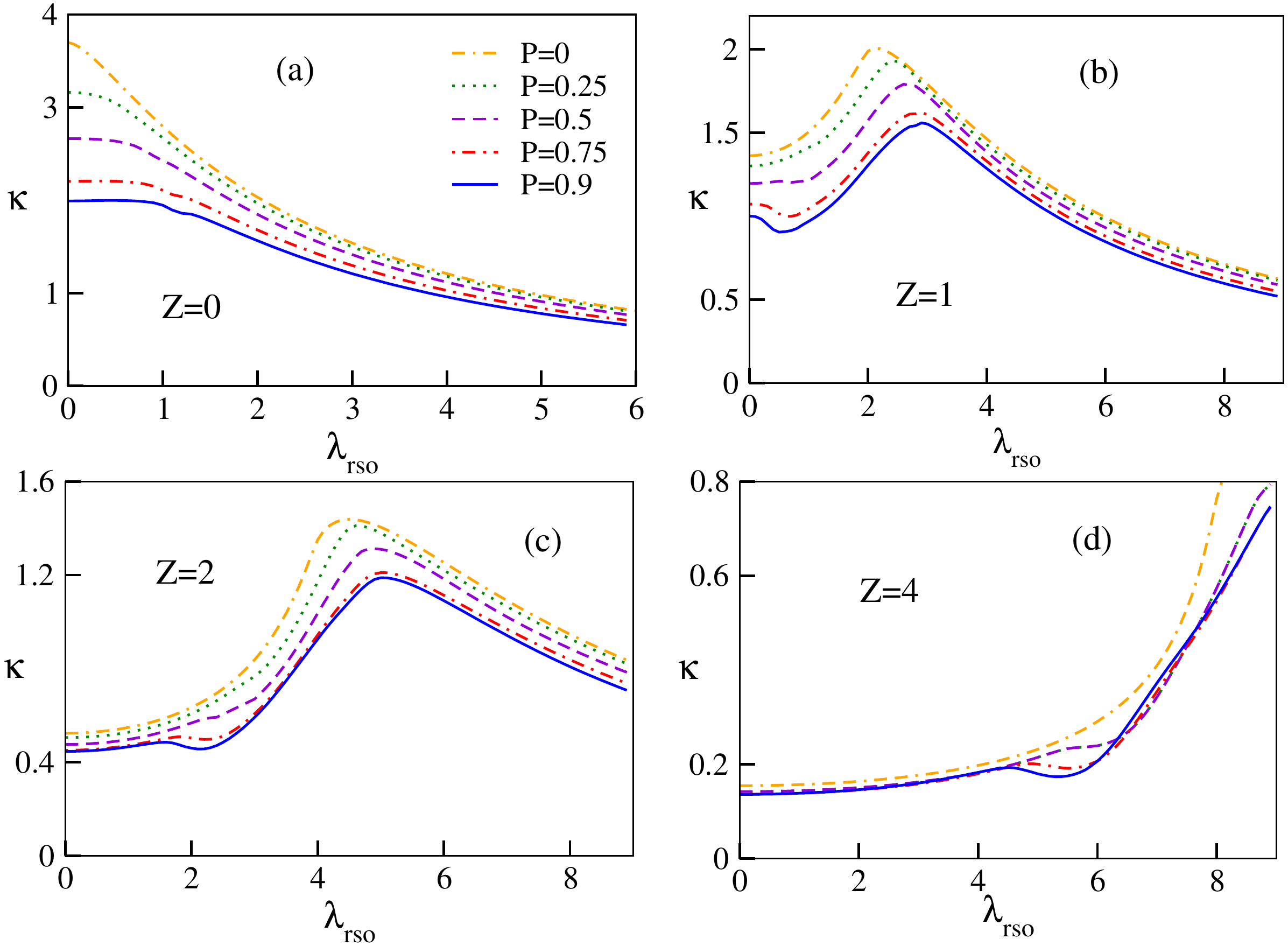}
\caption{The behavior of thermal conductance ($\kappa$), in unit of $k_B^2/h$, is displayed as a function of RSOI strength $\lambda_{rso}$ 
for a particular temperature ($T/T_c=0.7$) and different values of the barrier strength ($Z$) and polarization ($P$) of the ferromagnet.}
\label{cond_R}
\end{center}
\end{figure}

For illustration, we present in Fig.~\ref{cond_R} the behavior of \tc as a function of the RSOI strength for a particular temperature 
$T/T_c=0.7$. Here, (a), (b), (c) and (d) represent different barrier strength, $Z=0$, $1$, $2$ and $3$, respectively. In absence of scalar 
barrier ($Z=0$), when we increase the RSOI strength \tc monotonically decreases irrespective of the polarization (see Fig.~\ref{cond_R}(a)). 
Ferromagnet-superconductor structure with $P=0$ is exactly equivalent to a conventional normal metal-superconductor 
junction~\cite{chandrasekhar2009} for which the sub-gapped contribution to $\kappa$ is zero. The reason is that within the sub-gapped regime 
total AR probability is exactly equal to $1$ \ie~$R_{\sigma}^h=1$ with the reflection probability $R_{\sigma}^e=0$ as $Z=0$~\cite{blonder}. 
Therefore, the sub-gap contribution to the \tc is zero for this particular case. This is also evident from Eq.(\ref{kappa_form}). Only 
quasi-particle tunneling contributes to the \tc above the gap. Nevertheless, in presence of finite polarization of the ferromagnet the 
sub-gap contribution is finite. 
With the inclusion of interfacial RSOI, the spin-flip counter-parts of both types of scattering processes, AR and tunneling as electron or 
hole like quasi-particles, occur at the interface in addition to the spin-conserving processes. The relative amplitudes corresponding to the spin-conserving and 
spin-flipping processes will be completely determined by the strength of RSOI. As we increase $\lambda_{rso}$ spin-flip counter part starts 
dominating. Hence, SAR becomes higher in magnitude with the enhancement of $\lambda_{rso}$. With the increase of SAR, $\kappa$ decreases. 
Spin-flip counter parts of NR will increase in this case but its effect on \tc is less dominating compared to AR. As a consequence we 
get monotonically decreasing behavior of $\kappa$. Now with a particular RSOI strength, increasing polarization means reduction of AR due 
to the minority spin sub-band. SAR probability increases keeping the total AR probability $1$ and maintaining the usual result for the 
sub-gapped energy regime whereas tunneling corresponding to the minor spin sub-band reduces. This effectively reduces $\kappa$ by considerable 
amount as during the evaluation of $\kappa$,  all the electrons corresponding to all the energy values are taken into account. 

Comparing the magnitudes of $\kappa$ as shown in four figures of Fig.~\ref{cond_R}, it is evident that introduction of finite barrier $Z$ 
effectively reduces TC, being independent of the value of RSOI strength, due to the finite NR from the barrier.
In presence of barrier when we increase RSOI, initially $\kappa$ does not show any appreciable change as long as $\lambda_{rso}$ is very 
small compared to $Z$. If we gradually increase RSOI strength, \tc exhibits non-monotonic behavior. $\kappa$ increases gradually, attains 
a maxima and then again decreases with the increase of $\lambda_{rso}$ as depicted in Fig.~\ref{cond_R}(b). This is one of the main results
of the present article. The reason behind this non-monotonic behavior can be attributed to the interplay of all the six scattering processes 
occurring at the interface of FS structure. With the increase of $\lambda_{rso}$, probabilities of all the spin-flip scattering processes 
increase. However, for a low barrier strength ($Z=1$) SNR cannot dominate $\kappa$ significantly compared to the spin-flip tunneling. 
Interplay of these processes results in enhancement of $\kappa$ with the rise of $\lambda_{rso}$ accordingly. For sufficiently higher 
$\lambda_{rso}$ all the spin-flip processes start dominating over the spin-conserving processes. After a certain enhancement of RSOI, the 
probability of spin-flip scattering processes do not dominate further. Instead the spin-conserving scattering probabilities decrease with 
the increase of $\lambda_{rso}$. As a consequence, TC decreases. Behavior of $\kappa$ with the change of polarization remains monotonically 
decreasing similar to the case before.

If we further increase $Z$, $\kappa$ maintains its non-monotonic behavior with the rise of RSOI strength. However, the maxima moves towards 
the higher value of $\lambda_{rso}$. To illustrate this, we refer to Fig.~\ref{cond_R}(c) and (d). For higher values of $Z$, magnitudes of 
$\kappa$ decreases as SNR and NR starts dominating. In this situation, to obtain the maxima of $\kappa$ we have to tune RSOI strength
accordingly. This results in shifting of the peaks of $\kappa$ towards higher $\lambda_{rso}$. 

Note that, we present our results of $\kappa$ as a function of $\lambda_{rso}$ for $T/T_c=0.7$. If we consider the low temperature regime, 
particularly lower than the cross-over value $T_{x}$, we expect similar non-monotonic behavior of the \tc with RSOI strength. There will
be only one change in the nature of $\kappa$. Following the discussion in the previous subsection, at a particular RSOI strength $\kappa$ 
will rise with the increase of $P$ when temperature is lower than the cross-over value. However, for very low temperature such as 
$T/T_c \sim 0.1$, the amount of change in magnitude of the \tc is vanishingly small.
\subsection{Seebeck coefficient}
In this sub-section we study the phenomenon of Seebeck effect which is a direct measure of the local thermoelectric response. The behavior 
of \sbk as a function of temperature and RSOI for different values of $P$ and $Z$ are presented. All the results, we have presented here, 
\begin{figure}[htb]
\begin{center}
\includegraphics[width=8.7cm,height=7.3cm]{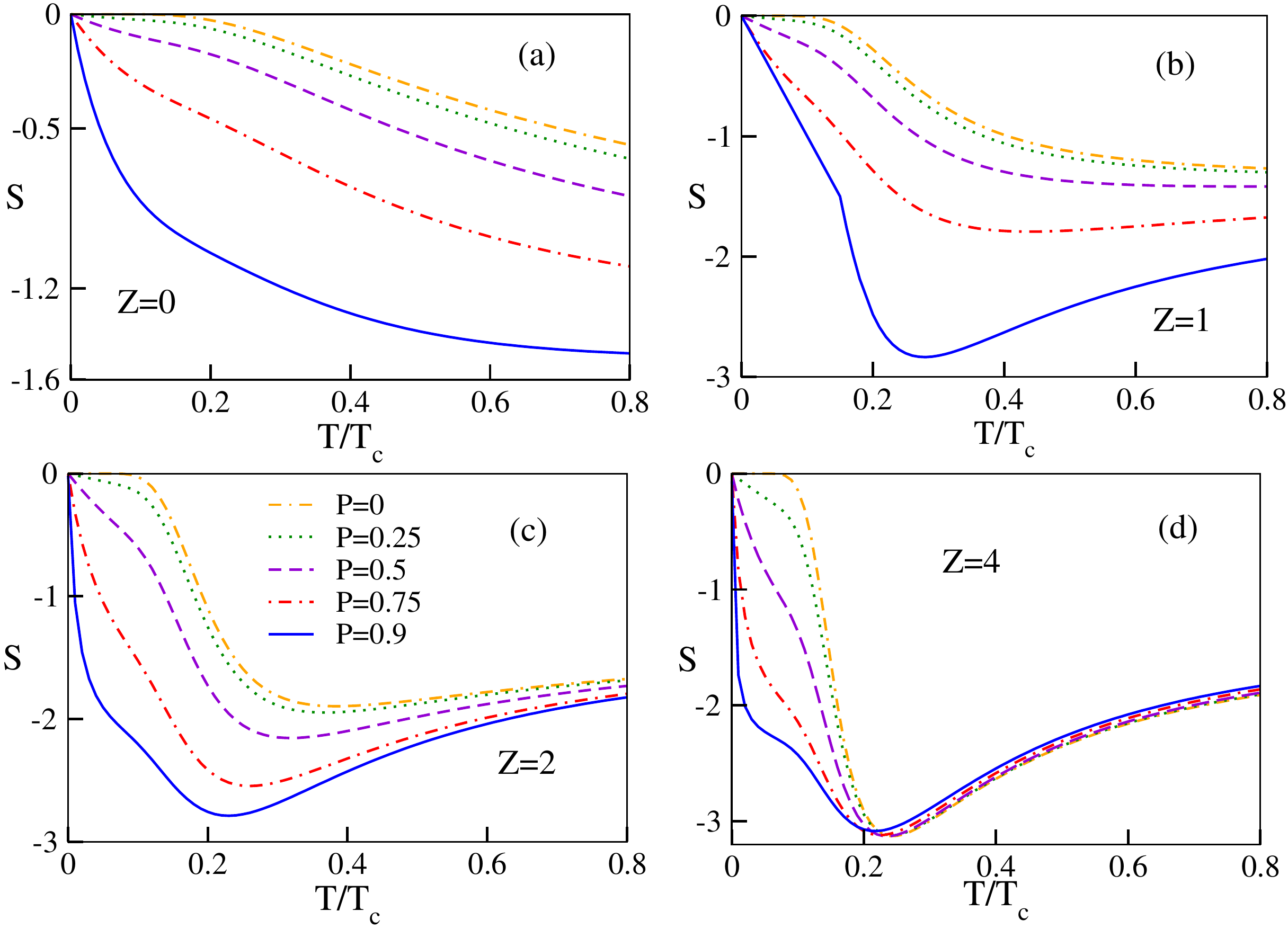}
\caption{(Color online) The behavior of Seebeck coefficient ($S$), in unit of $k_B/e$, is shown as a function of temperature ($T/T_c$) in 
absence of RSOI ($\lambda_{rso}=0$) for different values of the barrier strength ($Z$) and polarization ($P$) of the ferromagnet.}
\label{sbk_T}
\end{center}
\end{figure}
correspond to the particular orientation of the magnetization vector as mentioned in the previous sub-section. We divide our discussion 
in two different parts in order to highlight the effect of polarization of the ferromagnet and the effect of the interfacial RSOI on 
thermopower as follows.
\subsubsection{Effect of polarization and barrier in absence of RSOI}
In Fig.~\ref{sbk_T}[(a), (b), (c) and (d)], we show the behavior of $S$ as a function of $T/T_c$ corresponding to the scalar barrier potential 
$Z=0$, $1$, $2$ and $4$, respectively. In all of the four figures we observe that \sbk is negative throughout the window for all values of the 
polarization of the ferromagnet irrespective of the value of $Z$. It is solely due to the contributions arising from the electrons. So, we 
discuss only the magnitude of the \sbk throughout the rest of the manuscript. \sbk increases with $T/T_c$ in absence of the barrier \ie $Z=0$. 
This phenomenon is true for all values of the polarization $P$. On the other hand, for a particular value of temperature, it increases with the enhancement
of $P$. We can explain this phenomenon as follows. For $Z=0$, the probability of NR is zero. With the increase of temperature, superconducting 
gap decreases minimizing the phenomenon of AR to occur. This results in enhancement of transmission of energy with temperature. It is evident 
from the expression of \sbk (see Eq.(\ref{sbk_exp})). Hence, we can say that reduction of $R^h_{\sigma}$ causes enhancement of the magnitude of 
the numerator in Eq.(\ref{sbk_exp}). The latter essentially describes the thermal coefficient. At the same time denominator or the electronic 
conductance decreases. This effectively enhances \sbk magnitude. In fact, for the sub-gapped energy regime we always have zero thermal 
coefficient resulting in vanishing contribution to the \sbk. There are finite contributions arising only from the energy regime above the gap.

Now we consider barrier at the junction ($Z \ne 0$) and investigate how the thermopower behaves with the change of polarization in presence 
of finite barrier. In presence of finite $Z$, thermopower increases monotonically for low values of $P$. The amount of enhancement is higher 
for the higher values of $P$. However there is a smooth transition from monotonic to non-monotonic behavior of \sbk with temperature as the 
polarization tends to the half-metallic limit \ie $P=0.9$. It initially rises with $T/T_c$ and then decreases in high temperature regime with 
an extremum in the Seebeck profile. According to the definition we can describe this feature of \sbk in terms of $\alpha$ and $G$. In presence 
of finite $Z$, NR reduces both $\alpha$ and $G$. Now change in \sbk occurs depending on their relative magnitudes. As a consequence, $S$ 
increases in presence of finite $Z$ in comparison to the case for $Z=0$ (comparing Figs.~\ref{sbk_T}(a) and (b)). Moreover, there is a tendency 
of saturation of the thermopower magnitude in the higher temperature regime irrespective of the values of $P$. For $Z=1$, it is more clear for 
higher values of $P$. For sufficiently higher temperature the gap reduces significantly resulting in enhancement of conductance. However, for 
higher polarization, $G$ does not increase with $T/T_c$ by appreciable amount after a certain temperature due to the absence of minority spin 
band. As a consequence, \sbk decreases resulting in a non-monotonic behavior especially for higher polarization. Such non-monotonic nature is 
much more pronounced for higher values of $Z$ (see Fig.~\ref{sbk_T}(c) and (d)). Additionally, the saturation regime is achieved faster for higher $Z$ 
with the enhanced NR. For higher barrier strength, \sbk changes by very small amount with $Z$ for a particular $P$ and $T/T_c$ due to the 
suppression of all other scattering processes by NR.

Note that, our results does not imply that the \sbk approaches unity in the limit of zero RSOI and zero polarization. All these values are as small as in agreement with those found in the existing literature. We present a quantitative comparison in Appendix~\ref{appndx3}.
\begin{figure}[htb]
\begin{center}
\includegraphics[width=8.7cm,height=7.2cm]{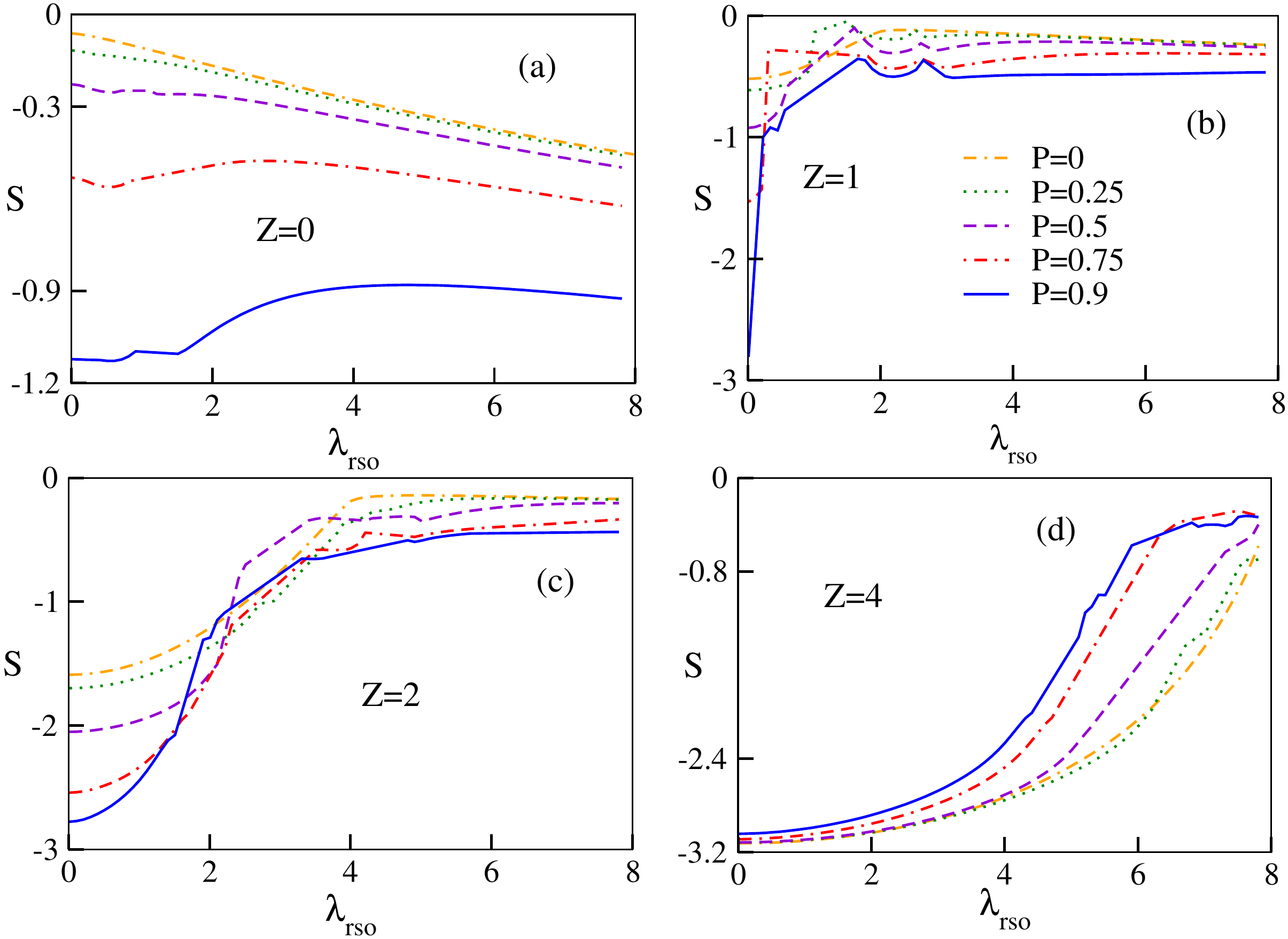}
\caption{(Color online) The behavior of Seebeck coefficient ($S$), in unit of $k_B/e$, at temperature $T/T_c=0.5$, is shown as a function of RSOI 
strength ($\lambda_{rso}$) for different values of barrier strength ($Z$) and polarization ($P$) of the ferromagnet.}
\label{sbk_R}
\end{center}
\end{figure}

\subsubsection{Interplay of polarization, barrier and interfacial RSOI}
In this sub-subsection, we investigate what happens to \sbk when RSOI is taken into account at the interfacial layer, both in absence and presence 
of finite scalar barrier at the junction. To understand the role of RSOI, we illustrate the behavior of \sbk ($S$) with respect to RSOI ($\lambda_{rso}$) 
for a fixed value of temperature $T/T_c=0.5$ in Fig.~\ref{sbk_R}[(a), (b), (c) and (d)] for four different values of the barrier strength 
as before. For $Z=0$, behavior of \sbk with $\lambda_{rso}$ are different for different $P$. For zero or lower values of polarization, 
it rises with the increase of RSOI strength. The rate of increase, however, changes when $P$ becomes high. There is a transition from 
increasing to decreasing nature of $S$ vs. $\lambda_{rso}$ curves with the increase of polarization. As soon as we incorporate $Z$, 
the situation changes dramatically. \sbk sharply falls with the temperature in presence of barrier strength due to the large boosts of NR 
and SNR scattering phenomenon. The rate of decrease of \sbk becomes lower with the increase of barrier strength. With very high RSOI it becomes
saturated with the saturation region shifting towards the higher RSOI strength as one increases the barrier strength $Z$.  This is clear by 
comparing Figs.~\ref{sbk_R}[(b)-(d)]. Most interestingly, 
\begin{figure}[htb]
\begin{center}
\includegraphics[width=8.7cm,height=7.5cm]{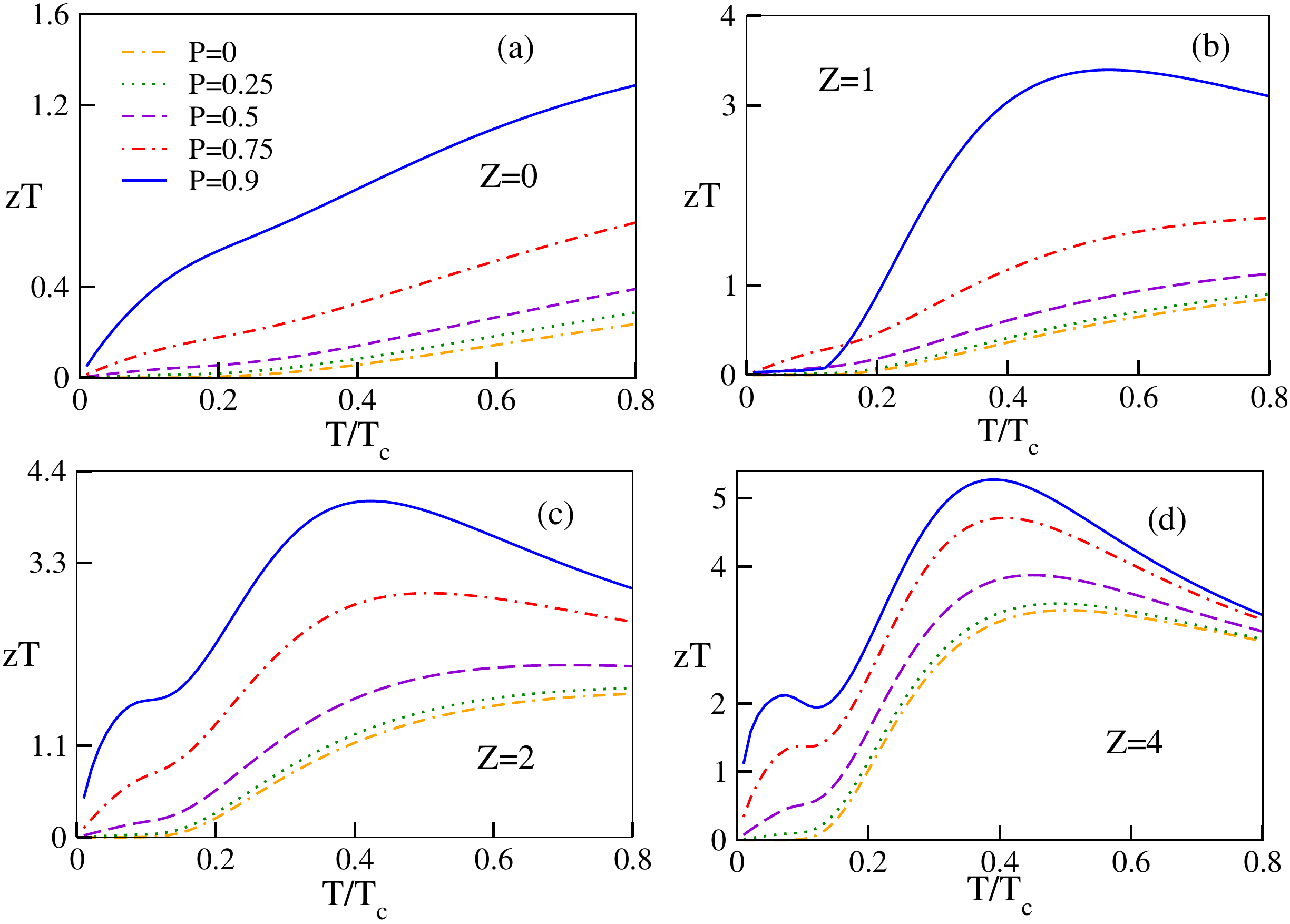}
\caption{(Color online) The feature of Figure of merit ($zT$) is depicted as a function of temperature ($T/T_c$) in absence of RSOI 
($\lambda_{rso}=0$) for different values of barrier strength ($Z$) and polarization ($P$) of the ferromagnet.}
\label{fom_T}
\end{center}
\end{figure}
we observe that for finite and low value of $Z$, \sbk always increases with polarization for all values of $\lambda_{rso}$. On the contrary, 
for higher $Z$ (\eg $Z=4$) we obtain exactly opposite scenario where $S$ decreases with the increase of $P$. There is a transition of the 
behavior of the \sbk with the polarization at some certain strength of $Z$. However, it is always possible to obtain large thermopower in 
presence of low but finite RSOI strength for all values of polarization. Note that, the behavior of $S$ with temperature remains quite
similar even in presence of fixed RSOI. For detailed discussion with more plots see Appendix~\ref{appndx2}.

Throughout our manuscript, we have focused on the low temperature regime for which our results are well justified. For higher temperature, close to or above $T_c$, 
one has to take the microscopic details of the self-consistency of the superconducting gap parameter into account which we have neglected in our toy model. 
\subsection{Figure of merit ($zT$)}
In order to understand the efficiency of our FS junction as a thermoelectric material, we explore the behavior of \fom 
both in presence and absence of RSOI at the interface. 

\subsubsection{Effect of polarization and barrier in absence of RSOI}
We discuss the role of polarization when we vary the FOM with temperature in absence of RSOI at the junction \ie $\lambda_{rso}=0$ as follows. 
In Fig.~\ref{fom_T} we display the variation of $zT$ with respect to $T/T_c$ for various polarizations and barrier strengths. Here, (a), (b), 
(c) and (d) represent the same parameter values as mentioned in the previous cases. The value of $zT$ is very small for $P=0$ and in the 
absence of barrier (see Fig.~\ref{fom_T}(a)). $zT$ is exactly zero if we consider only energy values within the subgapped regime. 
However, it increases with the increase of polarization as well as temperature. We can obtain FOM $\sim1.3$ when the polarization of the ferromagnet 
tends to half-metallic limit ($P=0.9$). The rate of increase of $zT$ with $P$, for a particular temperature, rises with the increase of polarization of the ferromagnet. 
In general, FOM can be defined in terms of the three parameters namely thermal coefficient, \sbk and electronic conductance (see Eq.(\ref{merit})). 
The nature of \fomd, whether it is increasing or decreasing, completely depends on their relative magnitudes. Similar enhancement of $zT$ has also been 
proposed in literature by tuning exchange field~\cite{machon}. Search for FOM more than $1$, in order to have an efficient thermoelectric material, has been always 
remained in the focus of material science. In our case FOM can be enhanced not only by tuning the polarization $P$ but also setting 
\begin{figure}[htb]
\begin{center}
\includegraphics[width=8.7cm,height=7.5cm]{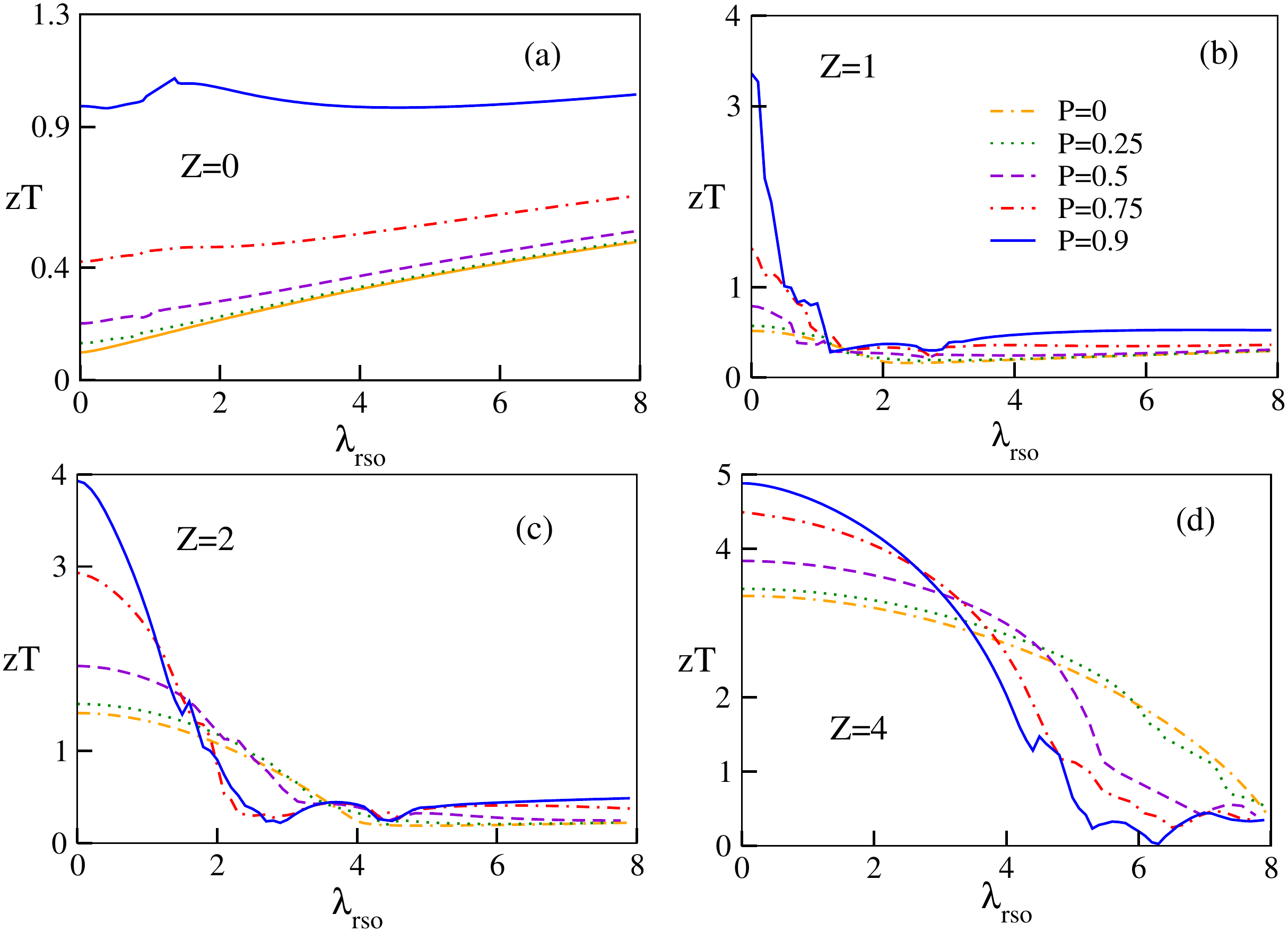}
\caption{(Color online) The behavior of Figure of merit ($zT$) is illustrated as a function of RSOI strength ($\lambda_{rso}$) at $T/T_c=0.5$ 
for different values of the barrier strength ($Z$) and polarization ($P$) of the ferromagnet.}
\label{fom_R}
\end{center}
\end{figure}
finite barrier potential at the junction as shown in Figs.~\ref{fom_T}[(b)-(d)]. For higher values of $Z$, the behavior of $zT$ becomes 
non-monotonic. It initially rises with the temperature and after a certain temperature it decreases. Remarkably, unlike \sbk, the point of 
maxima does not shift by appreciable amount with the change of barrier strength. More over, the sensitivity to the polarization reduces 
for higher strength of the barrier potential. 

Note that, in the limit of $\lambda_{rso}=0$, $Z=0$ and $P=0$, a comparison of our calculated values of \fom with the existing literature can be found 
in Appendix~\ref{appndx3}.

\subsubsection{Interplay of polarization, barrier and interfacial RSOI}
In presence of finite interfacial RSOI, the behavior of $zT$ with temperature is quite similar to that in absence of Rashba. A discussion on 
this issue can be found in Appendix~\ref{appndx2}. However, the magnitude of $zT$ depends on the strength of RSOI both in absence and presence 
of $Z$. We demonstrate the behavior of $zT$ as a function of $\lambda_{rso}$ in Fig.~\ref{fom_R} where (a)-(d) correspond to the same values of 
$Z$ as in the previous figures. 

In absence of any barrier ($Z=0$) and for low polarization of the ferromagnet, $zT$ rises with the enhancement of $\lambda_{rso}$ almost linearly
as shown in Fig.~\ref{fom_R}(a). Such linear behavior changes when the polarization of the ferromagnet is considered in the limit of half-metal, 
\ie $P=0.9$. For a particular choice of $\lambda_{rso}$, \fom increases with the increase of polarization too as in absence of RSOI (see 
Appendix~\ref{appndx2} for more details). This scenario reverses its character in presence of low $Z$. It is clear from Fig.~\ref{fom_R}(b). Under such 
circumstances, $zT$ falls off rapidly with the increase of $\lambda_{rso}$. As mentioned earlier, presence of RSOI induces 
spin-flip scattering process which reduces the \sbk and enhances \tc at the same time. The electronic conductance increases by 
small amount as shown in Ref.~\onlinecite{hogl}. When we further increase $Z$, the behavior of $zT$ changes slowly with the RSOI strength. Depending 
on the value of $Z$, (see Figs.~\ref{fom_R}(c) and (d)) \fom can even reach the value $\sim 5$ by using half-metal for weak or moderate RSOI. After a 
critical Rashba strength it falls off rapidly and becomes vanishingly small. However, the critical value of Rashba strength translates towards the 
higher regime with the increase of $Z$.

The main result of our article is that we have obtained large values of the thermoelectric coefficient in presence of a scalar barrier and/or 
Rashba spin-orbit interaction (RSOI) at the interface when polarization of the ferromagnet tends towards the half-metallic limit. Note that, 
the maximum value of $zT$ appears at a temperature scale which is well below the critical temperature of the superconductor. The physical reason 
behind such large values of the thermoelectric coefficients can be attributed to the interplay between scalar barrier strength, RSOI strength at 
the interface, polarization of the ferromagnet and quasiparticle tunneling, either spin-conserving or spin-flipping, through the interfacial 
barrier. Enhancement of thermoelectric coefficient due to the quasiparticle tunneling has been already established in literature but they 
involve either spin-split superconductor or spin-active interface~\cite{machon} corresponding to spin-dependent phase-shift~\cite{seviour2000giant}. 
In our case, the significant enhancement of the thermoelectric effect lies in the quasi-particle tunneling through the barrier breaking the 
spin-symmetry by Rashba spin-orbit interaction.

Therefore, we can say that \fom can be enhanced by increasing any one of the parameters, \ie polarization, barrier strength and temperature keeping 
the other two fixed both in absence and presence of RSOI at the interfacial layer of our FS junction. Note that, all these phenomena are true for weak or moderate Rashba interaction.

\section{Summary and Conclusions} \label{conclu}
To summarise, in this article, we have explored thermoelectric properties of a ferromagnet-superconductor junction driven by a thermal gradient. 
At the interfacial layer, we have incorporated RSOI along with scalar potential barrier. We have analyzed our results for the thermal conductance,
Seebeck coefficient and figure of merit as a function of temperature, polarization of the ferromagnet and RSOI strength. Thermal conductance 
$\kappa$ exhibits exponential rise with temperature following the fully developed gap feature. We have observed a cross-over temperature $T_{x}$ 
separating the two different regimes corresponding to opposite behaviors of $\kappa$ with the polarization of the ferromagnet. Below the 
cross-over temperature, $\kappa$ increases with the increase of polarization, whereas for all the temperatures above the cross-over temperature 
it decreases with polarization. This phenomenon is true both in absence and presence of barrier strength. With the increase of barrier strength 
the cross-over temperature moves towards the higher temperature value. Inclusion of interfacial Rashba spin-orbit field causes reduction of 
thermal conductance due to the appearance of spin-flipped SAR process, in absence of any barrier. On the other hand, interfacial RSOI can enhance 
the \tc in presence of finite barrier due to the interplay of both of them. We have obtained a non-monotonic behavior of the thermal conductance 
$\kappa$ as we vary the interfacial RSOI strength. The maxima of $\kappa$ moves towards the critical temperature $T_c$ with the enhancement of 
barrier strength $Z$. Throughout the manuscript we have considered only the electronic contribution to the \tc neglecting the contribution due 
to phonons which is a valid approximation for very small temperature gradient.

We have investigated the Seebeck coefficient in absence and presence of interfacial RSOI and scalar barrier. It can be enhanced by using 
ferromagnet with higher polarization and by increasing the temperature (below the critical temperature $T_{c}$) as well. A non-linear behavior 
of the \sbk with the temperature can be obtained in presence of barrier potential at the junction. Whereas presence of RSOI can reduce or enhance 
it depending on the barrier strength. For low barrier strength, \sbk increases with the increase of polarization. However, an exactly opposite 
behavior is observed when barrier strength becomes high. 

To quantify the thermoelectric power efficiency of the FS structure, we compute the figure of merit $zT$ both in presence and absence 
of RSOI and scalar potential barrier. We show that the \fom can be enhanced to more than $1$ by setting the polarization of the ferromagnet very 
high (towards the half-metallic limit). It can further be enhanced by considering finite potential barrier at the junction. In particular, the 
higher the barrier strength the higher the \fom turns out to be while inclusion of RSOI can reduce it for low barrier strength. For
higher barrier strength, value of $zT$ can again be more than $1$ ($zT\sim 4-5$) depending on the polarization of the ferromagnet. These phenomena are 
valid for weak Rashba spin-orbit interaction. Presence of strong RSOI can highly reduce it irrespective of the strength of the barrier potential.

As far as the practical realization of our model is concerned, it may be possible to fabricate such a FS hybrid structure by growing thin layers 
of a spin singlet superconductor (e.g. $\rm Nb$ material) and a ferromagnetic insulator (EuO)~\cite{matthias1961ferromagnetic,tokuyasu1988proximity} 
on top of each other. To design the interfacial layer, responsible for the spin-orbit field, one can use a thin layer of 
zinc-blende semiconductor~\cite{deng2012anomalous}. An additional gate 
voltage can be implemented to create the scalar potential barrier at the interface. The magnetization vector of the ferromagnet can be rotated 
by some external magnetic field. Additional effects of such external field can be avoided by using dysprosium magnets~\cite{hogl}. 
From an experimental point of view, the polarization of the ferromagnet can be a more controllable parameter than the interfacial 
RSOI strength. For a typical $s$-wave superconductor, for \eg $\rm Nb$ with $T_{c}=9.3~\rm{K}~(\Delta_{s}\sim 3 \rm~{meV})$, to attain $zT\sim 4.5$ 
one needs a ferromagnetic exchange coupling $\Delta_{xc}\sim 0.6 \rm~{eV}$~\cite{Steeneken2002,Tiffany2004}, temperature $T\sim 4 \rm~{K}$, 
interfacial Rashba parameter $\sim 60-80 \rm~{meV \cdot \AA{}}$~\cite{manchon3}. A scalar potential barrier with height $V\sim 0.5 \rm~{eV}$
and width $d \sim 2-4 \rm~{nm}$~\cite{Huang2011} is also required. 

In conclusion, we have shown the possibility of obtaining enhanced thermoelectric properties of a FS hybrid structure in presence of finite barrier 
and interfacial RSOI. In the context of thermal transport, the idea of exploring interfacial RSOI in FS junction is unique. The enhancement of the 
thermoelectric properties caused by the interplay of RSOI along with potential barrier at the interface of the heterostructure and the polarization 
of the ferromagnet elevate the potential of the FS structure as a thermoelectric. The proposed set-up, within the experimentally achievable parameter 
regime, may be utilized in thermoelectric devices. 


\vspace{.2cm}
\acknowledgments{PD thanks Department of Science and Technology (DST), India for the financial support through SERB NPDF 
(File no. PDF/2016/001178). We acknowledge S. D. Mahanti for helpful discussions and encouragement. 
AMJ also thanks DST, India for financial support.}

\begin{appendix}
\section{Outline of derivation of the thermoelectric coefficients}
\label{appndx1}

In presence of either a small bias or small temperature gradient across the junction, the linear response for the charge 
and heat currents can be expressed following the Onsager matrix equation~\cite{onsager1931reciprocal1,onsager1931reciprocal2} given as,
\bea
\left( \begin{array}{c}
I \\
I_Q 
\end{array}\right)
&=&
\left( \begin{array}{c c }
L_{11} & L_{12} \\
L_{12} & L_{22} T
\end{array}\right)
\left( \begin{array}{c}
V \\
\Delta T/T 
\end{array}\right)\ 
\eea
where $L_{11}$ is the electrical conductance describing the charge current flowing due to the applied bias, $L_{22}$ is the thermal conductance
describing the heat flow driven by the temperature gradient and $L_{12}$ is the thermoelectric coefficient corresponding to heat flow due to the 
bias or charge current due to temperature gradient. 

Following BTK approach~\cite{blonder,wysokinski2012thermoelectric}, in presence of bias $V$ and temperature gradient $\delta T$, 
we can express the net current flowing through the FS junction as,
\bea
I_{\rm FS}&=&2 N(0) e v_F \mathcal{A} \int\limits_{-\infty}^{\infty} \Big[ \{1-R^e(E)\} f_0(E-eV,T+\delta T/2)\Big.\non\\
&&~~~~~~~~~~~~~~~~~~~~~~~~ -R^h(E) f_0 (E+eV,T+\delta T/2) \Big. \non \\
&& -\{1-R^e(E)-R^h(E)\}f_0(E,T-\delta T/2) \Big]dE \ 
\label{btk}
\eea 
where $N(0)$ is the density of states at the Fermi level, $e$ is the electronic charge, $v_F$ is the Fermi velocity and $\mathcal{A}$ 
is the area of contact. For very small applied bias voltage and temperature gradient, we can expand the Fermi distribution functions 
in Taylor series as follows,
\bea
f_0(E-eV,T+\delta T/2)&=&f_0 (E,T)-eV \frac{\partial f_0}{\partial E} +\frac{\delta T}{2} \frac{\partial f_0}{\partial T} \non \\
f_0(E+eV,T+\delta T/2)&=&f_0 (E,T)+eV \frac{\partial f_0}{\partial E} +\frac{\delta T}{2} \frac{\partial f_0}{\partial T} \non \\
f_0(E,T-\delta T/2)&=&f_0 (E,T)-\frac{\delta T}{2} \frac{\partial f_0}{\partial T}.
\eea
Substituting the expressions of the distribution functions, we simplify Eq.(\ref{btk}) for finite but small (linear response regime) temperature gradient 
in absence of bias as,
\bea
I_{\rm FS}^{\prime} &=& 2 N(0) e v_F \mathcal{A} \int \limits_{-\infty}^{\infty} \Big[1-R^e(E)-R^h(E)\Big] \delta T \frac{\partial f_0}{\partial T} dE \ .\non \\
\eea
Therefore the current flow through the junction per unit temperature difference is obtained as
\bea
\frac{I_{\rm FS}^{\prime}}{I_0}=\int \limits_{0}^{\infty}\frac{\Big[ 1-R^e(E)-R^h(E) \Big]}{k_B T^2 \cosh^2(\frac{E-E_F}{2 k_B T})} \delta T (E-E_F)~dE \  \non\\
\label{IFS2}
\eea
where the normalization constant is given by $I_0=N(0)e v_F\mathcal{A}$. Similarly, if we consider only the bias instead of the temperature gradient
we obtain the expression for current as,
\bea
\frac{I_{\rm FS}^{\prime\prime}}{I_0}=\int \limits_{0}^{\infty}\frac{\Big[ 1-R^e(E)+R^h(E) \Big]}{k_B T^2 \cosh^2(\frac{E-E_F}{2 k_B T})} eV ~dE \ . \non\\
\label{IFS3}
\eea
From Eq.(\ref{IFS3}) we obtain the mathematical expression of electrical conductance or the well-known BTK formula for the electrical conductance~\cite{blonder} 
in the linear response regime as,
\bea
L_{11}= \int \limits_{0}^{\infty}\frac{\Big[1-R^e(E)+R^h(E) \Big]}{k_B T \cosh^2(\frac{E-E_F}{2 k_B T})} dE.
\eea
After performing the wave vector integration we can get back Eq.(\ref{intG}) for the electrical conductance $G$ normalized by the Sharvin conductance $G_0$ 
(in unit of $e^2/h$) for a perfect contact as mentioned in Ref.~\onlinecite{hogl}.

On the other hand, thermal conductance can be defined as the heat current flowing across the junction per unit temperature difference and given 
by~\cite{xu2014enhanced,linder2016},
\bea
L_{22}=\int \limits_{0}^{\infty}\frac{\Big[1-R^e(E)-R^h(E) \Big]}{k_B T^2 \cosh^2(\frac{E-E_F}{2 k_B T})} \frac{(E-E_F)^2}{k_B e^2} dE.
\eea
In our case additional summation for the electron spin as well as an integration over the wave vector parallel to the interface have to 
be taken into account following Ref.~\onlinecite{hogl} as written in the expression of normalized thermal conductance $\kappa$ in 
Eq.(\ref{kappa_form}). The normalization constant $k_0$ is $G_0 k_B^2 T/e^2$ ($\equiv k_B^2 T/h$).

\begin{figure}[htb]
\begin{center}
\includegraphics[width=8.7cm,height=7.cm]{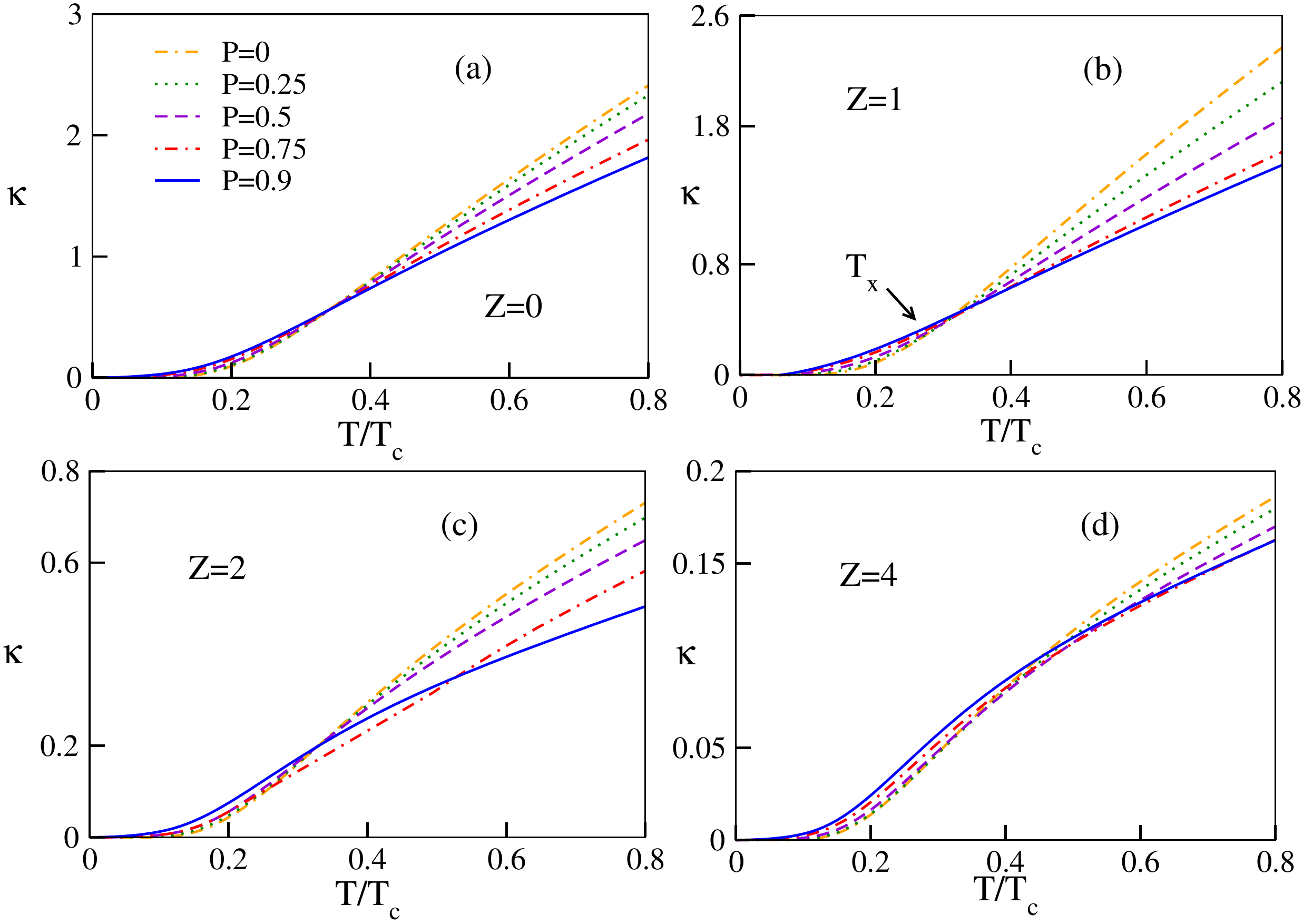}
\caption{(Color online) The variation of thermal conductance ($\kappa$), in unit of $k_B^2/h$, with respect to temperature ($T/T_c$) 
is depicted in presence of RSOI ($\lambda_{rso}=2$) for different values of the barrier strength ($Z$) and polarization ($P$) of the 
ferromagnet.}
\label{rashbacond_T}
\end{center}
\end{figure}
Using Eq.(\ref{IFS2}) we can also write the expression for $L_{12}$ as,
\bea
L_{12}=\int \limits_{0}^{\infty}\frac{\Big[1-R^e(E)-R^h(E) \Big]}{k_B T \cosh^2(\frac{E-E_F}{2 k_B T})} \frac{(E-E_F)}{k_B e} dE \ .
\eea 
Following the route of electrical and thermal conductance, we perform the summation over spin and integration over wave vector to obtain 
\begin{figure}[htb]
\begin{center}
\includegraphics[width=8.7cm,height=7.2cm]{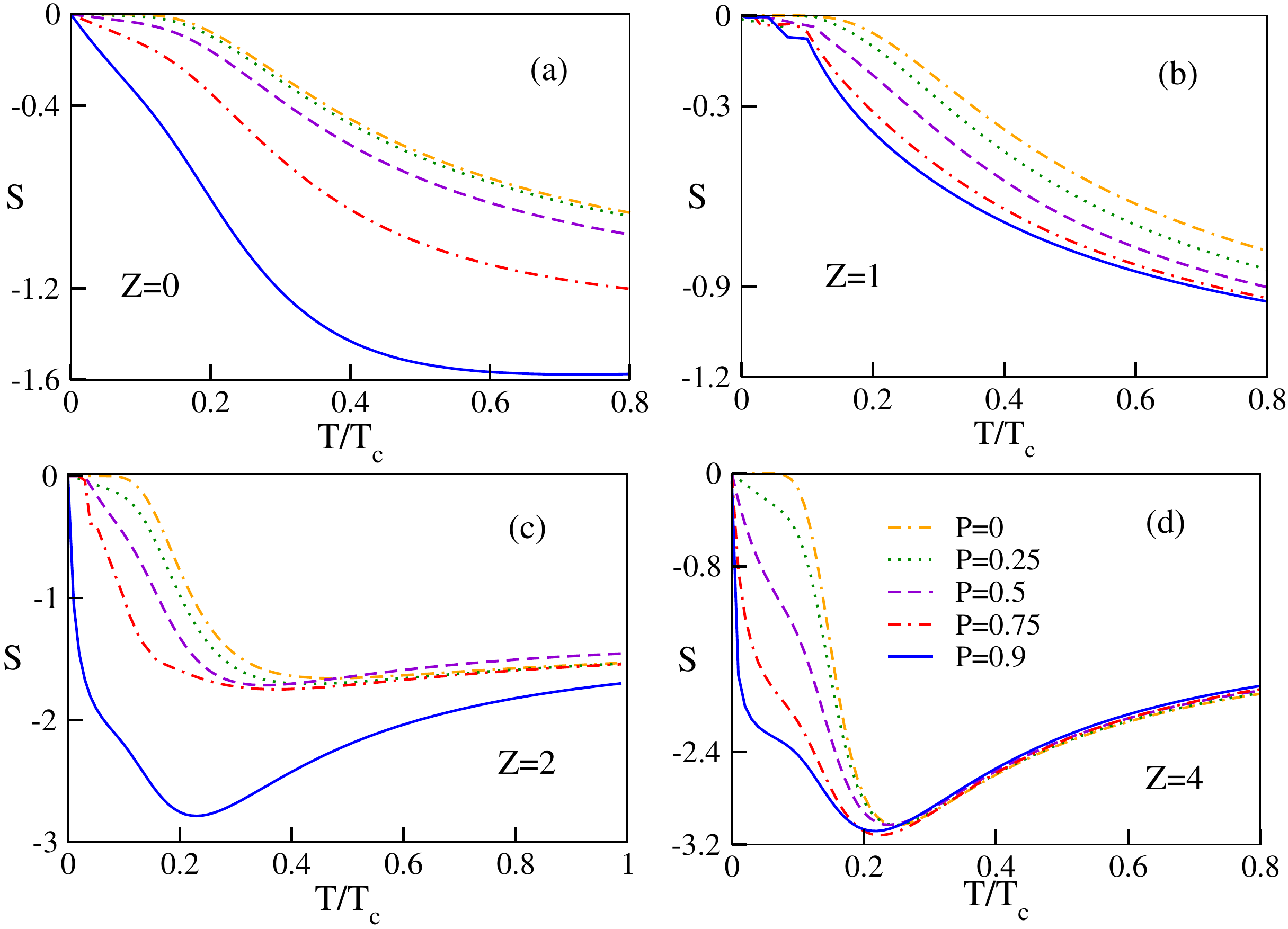}
\caption{(Color online) The variation of Seebeck coefficient ($S$), in unit of $k_B/e$, with respect to temperature ($T/T_c$) is displayed 
in presence of RSOI ($\lambda_{rso}=2$) for different values of barrier strength ($Z$) and polarization ($P$) of the ferromagnet.}
\label{rashbasbk_T}
\end{center}
\end{figure}
the exact form of the thermoelectric coefficient as provided in Eq.(\ref{alphaform}).

Thus, from Onsager matrix equation we write
\beq
I=L_{11} V + L_{12}\Delta T/T.
\eeq
In open circuit condition ($I=0$), the induced voltage per unit temperature gradient or the so-called Seebeck coefficient can be found 
in terms of $L_{12}$ and $L_{22}$ 
\bea
S=-\frac{1}{T} \frac{L_{12}}{L_{11}}.
\label{seebeck_L12}
\eea
It is measured in the unit of $k_B/e$ (V/K). Note that, we can obtain Eq.~(\ref{sbk_exp}) from Eq.~(\ref{seebeck_L12}) by replacing 
$L_{12}$ by $\alpha$ and $L_{11}$ by $G$ as given in Eq.(\ref{sbk_exp}) in unit of $k_B/e$.

Hence, incorporation of the units of $G$ ($e^2/h$), $S$ ($k_B/e$), and $K$ ($k_B^2 T /h$) will make the figure of \
merit $zT$ dimensionless.

\section{Behavior of the thermal conductance, Seebeck coefficient and figure of merit with temperature at fixed RSOI}
\label{appndx2}

In Fig.~\ref{rashbacond_T}, we show the variation of the \tc as a function of temperature ($T/T_c$) for finite RSOI strength. Here, 
Fig.~\ref{rashbacond_T}[(a), (b), (c) and (d)] represent the scalar barrier strength $Z=0$, $1$, $2$ and $4$, respectively. Comparing 
Fig.~\ref{cond_T}(a) and Fig.~\ref{rashbacond_T}(a) we observe that in presence of finite RSOI, \tc decreases irrespective of the degree 
of polarization of the ferromagnet for $Z=0$. Now incorporation of finite barrier strength $Z$ makes $\kappa$ behave differently from the 
$Z=0$ 
\begin{figure}[htb]
\begin{center}
\includegraphics[width=8.7cm,height=7.2cm]{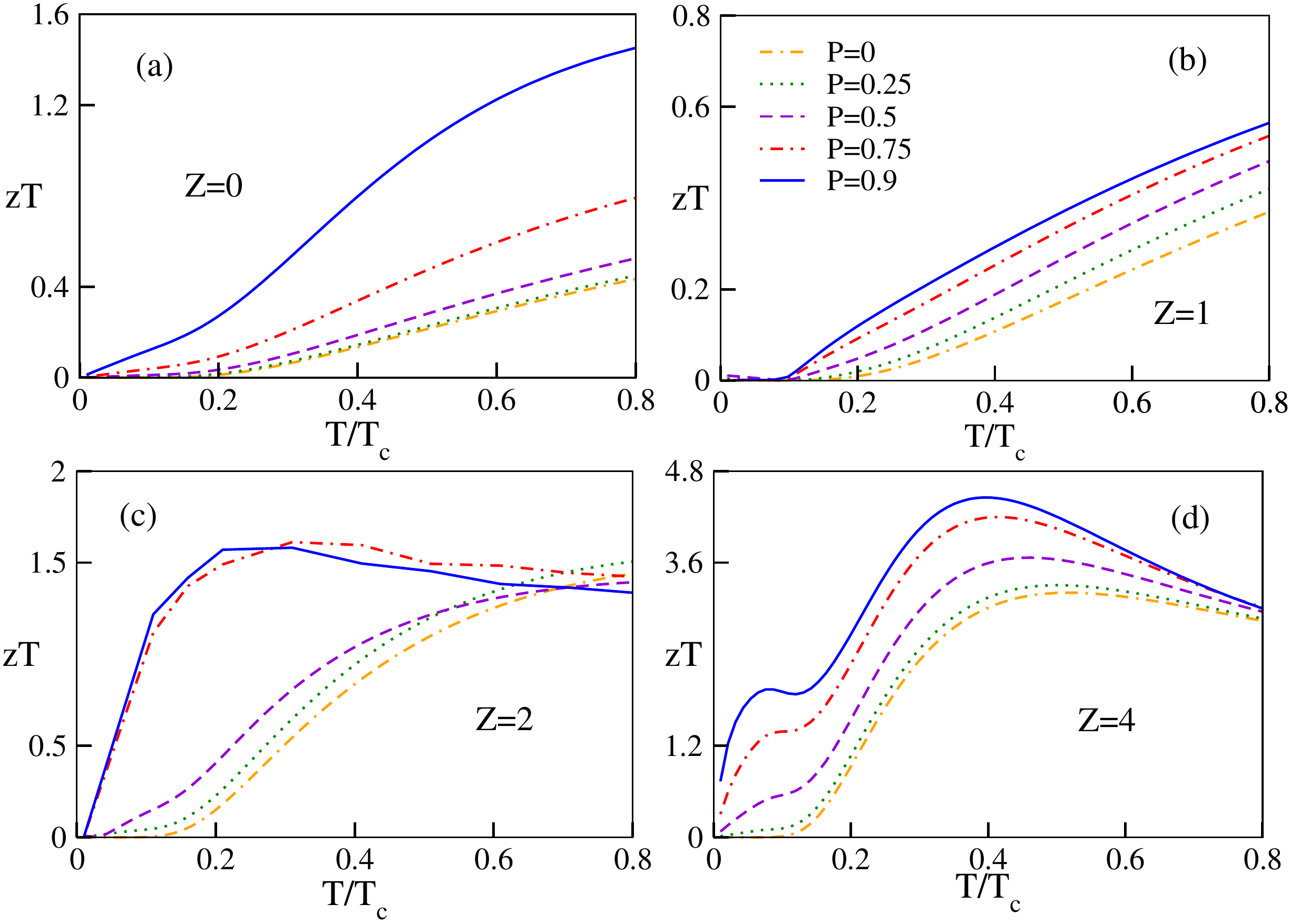}
\caption{(Color online) The variation of Figure of merit ($zT$) is shown as a function of temperature ($T/T_c$) in presence of finite RSOI 
($\lambda_{rso}=2$) for different values of barrier strength ($Z$) and polarization ($P$) of the ferromagnet.}
\label{rashbafom_T}
\end{center}
\end{figure}
case [see Fig.~\ref{rashbacond_T}(b)-(d)]. The scenario becomes much more interesting when we compare Fig.~\ref{cond_T}(b) and 
Fig.~\ref{rashbacond_T}(b). We notice that in presence of RSOI, behavior of $\kappa$ with RSOI is non-monotonic. It is clear from the fact 
that for $Z=1$, \tc increases while we introduce RSOI at the interface. On the other hand for higher values of $Z$ ($Z=2$ and $Z=4$), 
$\kappa$ decreases with the incorporation of RSOI. It is evident from the comparison of Fig.~\ref{cond_T} with Fig.~\ref{rashbacond_T} for 
all values of $Z$. Note that, the cross-over temperature separating the two opposite behaviors of the \tc with polarization also exists in 
presence of RSOI. Also the magnitude of $T_x$ shifts towards the high temperature regime with the enhancement of $Z$ similar to the case of 
in absence of barrier at the junction (see Fig.~\ref{rashbacond_T}(c) and (d)).

In Fig.~\ref{rashbasbk_T} we display the behavior of \sbk with $T/T_c$ in presence of RSOI for different values of $Z$ and $P$. Here, (a), 
(b), (c) and (d) represent $Z=0$, $1$, $2$ and $4$, respectively similar to the previous figures. Comparing Fig.~\ref{sbk_T}(a) and 
Fig.~\ref{rashbasbk_T}(a), it is clear that for $Z=0$ the nature of the curves for $S$ as a function of $T/T_c$, in absence and presence of RSOI, 
are quite similar to each other except the slopes. Note that, \sbk changes much faster with $T/T_c$ in presence of RSOI whereas for a particular
temperature, \sbk increases with the increase of polarization $P$. However, the behavior of \sbk changes significantly when we incorporate 
barrier strength at the FS interface. For low $Z$ (for \eg $Z=1$), \sbk gets reduced as soon as we incorporate RSOI at the interface (compare 
Figs.~\ref{sbk_T}(b) and \ref{rashbasbk_T}(b)). However, we recover the non-monotonic behavior of \sbk when $Z$ becomes high ($Z=2$ and $4$) 
as demonstrated in Figs.~\ref{rashbasbk_T}(c) and (d).

The behavior of \fom with temperature in presence of RSOI ($\lambda_{rso}=2$) is shown in Fig.~\ref{rashbafom_T}. In absence of any barrier 
at the interface, FOM increases with the rise of temperature, for a particular value of polarization. The behavior of $zT$ with temperature 
and also with polarization in presence of $\lambda_{rso}$ is very much similar to that in absence of RSOI which is clear by comparing 
Fig.~\ref{fom_T}(a) and Fig.~\ref{rashbafom_T}(a). However, the scenario changes drastically when we consider finite barrier at the junction. 
In presence of low $Z$, $zT$ decreases for all values of $P$ as shown in Fig.~\ref{rashbafom_T}(b). Moreover, the sensitivity of $zT$ to $P$ 
is very small for low $Z$. For higher barrier strength, exactly opposite situation occurs. The interplay of high potential barrier, polarization 
and RSOI (weak or moderate) causes enhancement of $zT$ with the increase of $Z$ (see Figs.~\ref{rashbafom_T}(c)-(d)) for the weak strength of 
RSOI ($\lambda_{rso}$).

\section{Comparison of our results with the existing literature for $\lambda_{rso}=0$, $P=0$ and $Z=0$}
\label{appndx3}
\end{appendix}
We compare the magnitudes of the \sbk and \fom with the results existing in the literature for the limit of zero interfacial RSOI ($\lambda_{rso}=0$) and zero exchange field ($P=0$).  Within this limit, we have found \sbk $\sim 0.3 \times 10^{-4}$ V/K at $T/T_c=0.5$ in absence of barrier ($Z=0$) (see orange curve of Fig.~\ref{sbk_T}(a)). In Ref.~\onlinecite{linder2016}, we observe that \sbk is zero for the same parameter regime. However, for other values of polarization and temperature, $S$ $\sim 10^{-4}$  V/K or $10^{-3}$ V/K depending on the values of $T$ and $P$ (see Ref.~\onlinecite{linder2016} for details). 
For $\lambda_{rso}=0$, $P=0$ and $Z=0$, our system basically converts to a normal metal-superconductor junction and we can compute our results directly from the BTK formalism using Eq.~(\ref{sbk_exp}). \sbk should be zero in this parameter regime only if we consider the energy values within the superconducting sub-gapped regime. Within this energy window, integration over energy gives rise to exactly zero thermopower. However, we have considered all the energy levels in our analysis and there are some finite contributions arising from the energy levels above the superconducting gap. As a consequence, we obtain low but finite $S$. Now, zero \sbk results in zero \fom (see Eq.~(\ref{merit})) as shown in Ref.~\onlinecite{linder2016}. However, due to the finite contributions arising from the energy levels above the 
superconducting gap, we have obtained very low but finite value of figure of merit. Note that, our system is not exactly equivalent to the system considered in Ref.~\onlinecite{linder2016}. Instead of two spin-split superconductors separated by a magnetic barrier, we have only one superconductor to form the junction with a ferromagnet. However, we have compared the values of $S$ and $zT$ with this reference only in absence of exchange interaction and found that the order of magnitude are well in agreement.


\begin{thebibliography}{58}%
\makeatletter
\providecommand \@ifxundefined [1]{%
 \@ifx{#1\undefined}
}%
\providecommand \@ifnum [1]{%
 \ifnum #1\expandafter \@firstoftwo
 \else \expandafter \@secondoftwo
 \fi
}%
\providecommand \@ifx [1]{%
 \ifx #1\expandafter \@firstoftwo
 \else \expandafter \@secondoftwo
 \fi
}%
\providecommand \natexlab [1]{#1}%
\providecommand \enquote  [1]{``#1''}%
\providecommand \bibnamefont  [1]{#1}%
\providecommand \bibfnamefont [1]{#1}%
\providecommand \citenamefont [1]{#1}%
\providecommand \href@noop [0]{\@secondoftwo}%
\providecommand \href [0]{\begingroup \@sanitize@url \@href}%
\providecommand \@href[1]{\@@startlink{#1}\@@href}%
\providecommand \@@href[1]{\endgroup#1\@@endlink}%
\providecommand \@sanitize@url [0]{\catcode `\\12\catcode `\$12\catcode
  `\&12\catcode `\#12\catcode `\^12\catcode `\_12\catcode `\%12\relax}%
\providecommand \@@startlink[1]{}%
\providecommand \@@endlink[0]{}%
\providecommand \url  [0]{\begingroup\@sanitize@url \@url }%
\providecommand \@url [1]{\endgroup\@href {#1}{\urlprefix }}%
\providecommand \urlprefix  [0]{URL }%
\providecommand \Eprint [0]{\href }%
\providecommand \doibase [0]{http://dx.doi.org/}%
\providecommand \selectlanguage [0]{\@gobble}%
\providecommand \bibinfo  [0]{\@secondoftwo}%
\providecommand \bibfield  [0]{\@secondoftwo}%
\providecommand \translation [1]{[#1]}%
\providecommand \BibitemOpen [0]{}%
\providecommand \bibitemStop [0]{}%
\providecommand \bibitemNoStop [0]{.\EOS\space}%
\providecommand \EOS [0]{\spacefactor3000\relax}%
\providecommand \BibitemShut  [1]{\csname bibitem#1\endcsname}%
\let\auto@bib@innerbib\@empty
\bibitem [{\citenamefont {Chandrasekhar}(2009)}]{chandrasekhar2009}%
  \BibitemOpen
  \bibfield  {author} {\bibinfo {author} {\bibfnamefont {V.}~\bibnamefont
  {Chandrasekhar}},\ }\href@noop {} {\bibfield  {journal} {\bibinfo  {journal}
  {Superconductor Science and Technology}\ }\textbf {\bibinfo {volume} {22}},\
  \bibinfo {pages} {083001} (\bibinfo {year} {2009})}\BibitemShut {NoStop}%
\bibitem [{\citenamefont {Machon}\ \emph {et~al.}(2014)\citenamefont {Machon},
  \citenamefont {Eschrig},\ and\ \citenamefont {Belzig}}]{machon}%
  \BibitemOpen
  \bibfield  {author} {\bibinfo {author} {\bibfnamefont {P.}~\bibnamefont
  {Machon}}, \bibinfo {author} {\bibfnamefont {M.}~\bibnamefont {Eschrig}}, \
  and\ \bibinfo {author} {\bibfnamefont {W.}~\bibnamefont {Belzig}},\
  }\href@noop {} {\bibfield  {journal} {\bibinfo  {journal} {New J. Phys.}\
  }\textbf {\bibinfo {volume} {16}},\ \bibinfo {pages} {073002} (\bibinfo
  {year} {2014})}\BibitemShut {NoStop}%
\bibitem [{\citenamefont {Ozaeta}\ \emph {et~al.}(2014)\citenamefont {Ozaeta},
  \citenamefont {Virtanen}, \citenamefont {Bergeret},\ and\ \citenamefont
  {Heikkil{\"a}}}]{ozaeta2014}%
  \BibitemOpen
  \bibfield  {author} {\bibinfo {author} {\bibfnamefont {A.}~\bibnamefont
  {Ozaeta}}, \bibinfo {author} {\bibfnamefont {P.}~\bibnamefont {Virtanen}},
  \bibinfo {author} {\bibfnamefont {F.}~\bibnamefont {Bergeret}}, \ and\
  \bibinfo {author} {\bibfnamefont {T.}~\bibnamefont {Heikkil{\"a}}},\
  }\href@noop {} {\bibfield  {journal} {\bibinfo  {journal} {Phys. Rev. Lett.}\
  }\textbf {\bibinfo {volume} {112}},\ \bibinfo {pages} {057001} (\bibinfo
  {year} {2014})}\BibitemShut {NoStop}%
\bibitem [{\citenamefont {Kalenkov}\ \emph {et~al.}(2012)\citenamefont
  {Kalenkov}, \citenamefont {Zaikin},\ and\ \citenamefont
  {Kuzmin}}]{kalenkov2012theory}%
  \BibitemOpen
  \bibfield  {author} {\bibinfo {author} {\bibfnamefont {M.~S.}\ \bibnamefont
  {Kalenkov}}, \bibinfo {author} {\bibfnamefont {A.~D.}\ \bibnamefont
  {Zaikin}}, \ and\ \bibinfo {author} {\bibfnamefont {L.~S.}\ \bibnamefont
  {Kuzmin}},\ }\href@noop {} {\bibfield  {journal} {\bibinfo  {journal} {Phys.
  Rev. Lett.}\ }\textbf {\bibinfo {volume} {109}},\ \bibinfo {pages} {147004}
  (\bibinfo {year} {2012})}\BibitemShut {NoStop}%
\bibitem [{\citenamefont {Machon}\ \emph {et~al.}(2013)\citenamefont {Machon},
  \citenamefont {Eschrig},\ and\ \citenamefont {Belzig}}]{machon2}%
  \BibitemOpen
  \bibfield  {author} {\bibinfo {author} {\bibfnamefont {P.}~\bibnamefont
  {Machon}}, \bibinfo {author} {\bibfnamefont {M.}~\bibnamefont {Eschrig}}, \
  and\ \bibinfo {author} {\bibfnamefont {W.}~\bibnamefont {Belzig}},\
  }\href@noop {} {\bibfield  {journal} {\bibinfo  {journal} {Phys. Rev. Lett.}\
  }\textbf {\bibinfo {volume} {110}},\ \bibinfo {pages} {047002} (\bibinfo
  {year} {2013})}\BibitemShut {NoStop}%
\bibitem [{\citenamefont {Kolenda}\ \emph
  {et~al.}(2016{\natexlab{a}})\citenamefont {Kolenda}, \citenamefont {Wolf},\
  and\ \citenamefont {Beckmann}}]{kolenda2}%
  \BibitemOpen
  \bibfield  {author} {\bibinfo {author} {\bibfnamefont {S.}~\bibnamefont
  {Kolenda}}, \bibinfo {author} {\bibfnamefont {M.~J.}\ \bibnamefont {Wolf}}, \
  and\ \bibinfo {author} {\bibfnamefont {D.}~\bibnamefont {Beckmann}},\
  }\href@noop {} {\bibfield  {journal} {\bibinfo  {journal} {Phys. Rev. Lett.}\
  }\textbf {\bibinfo {volume} {116}},\ \bibinfo {pages} {097001} (\bibinfo
  {year} {2016}{\natexlab{a}})}\BibitemShut {NoStop}%
\bibitem [{\citenamefont {Kolenda}\ \emph
  {et~al.}(2016{\natexlab{b}})\citenamefont {Kolenda}, \citenamefont {Machon},
  \citenamefont {Beckmann},\ and\ \citenamefont {Belzig}}]{kolenda2016}%
  \BibitemOpen
  \bibfield  {author} {\bibinfo {author} {\bibfnamefont {S.}~\bibnamefont
  {Kolenda}}, \bibinfo {author} {\bibfnamefont {P.}~\bibnamefont {Machon}},
  \bibinfo {author} {\bibfnamefont {D.}~\bibnamefont {Beckmann}}, \ and\
  \bibinfo {author} {\bibfnamefont {W.}~\bibnamefont {Belzig}},\ }\href@noop {}
  {\bibfield  {journal} {\bibinfo  {journal} {Beilstein J. Nanotechnol.}\
  }\textbf {\bibinfo {volume} {7}},\ \bibinfo {pages} {1579} (\bibinfo {year}
  {2016}{\natexlab{b}})}\BibitemShut {NoStop}%
\bibitem [{\citenamefont {Andreev}(1964{\natexlab{a}})}]{andreev1}%
  \BibitemOpen
  \bibfield  {author} {\bibinfo {author} {\bibfnamefont {A.~F.}\ \bibnamefont
  {Andreev}},\ }\href@noop {} {\bibfield  {journal} {\bibinfo  {journal} {Zh.
  Eksp. Teor. Fiz.}\ }\textbf {\bibinfo {volume} {46}},\ \bibinfo {pages}
  {1823} (\bibinfo {year} {1964}{\natexlab{a}})}\BibitemShut {NoStop}%
\bibitem [{\citenamefont {Abrikosov}\ and\ \citenamefont
  {Beknazarov}(1988)}]{abrikosov1988fundamentals}%
  \BibitemOpen
  \bibfield  {author} {\bibinfo {author} {\bibfnamefont {A.~A.}\ \bibnamefont
  {Abrikosov}}\ and\ \bibinfo {author} {\bibfnamefont {A.}~\bibnamefont
  {Beknazarov}},\ }\href@noop {} {\emph {\bibinfo {title} {Fundamentals of the
  Theory of Metals}}},\ Vol.~\bibinfo {volume} {1}\ (\bibinfo  {publisher}
  {North-Holland Amsterdam},\ \bibinfo {year} {1988})\BibitemShut {NoStop}%
\bibitem [{\citenamefont {Yokoyama}\ \emph {et~al.}(2008)\citenamefont
  {Yokoyama}, \citenamefont {Linder},\ and\ \citenamefont
  {Sudb{\o}}}]{yokoyama2008heat}%
  \BibitemOpen
  \bibfield  {author} {\bibinfo {author} {\bibfnamefont {T.}~\bibnamefont
  {Yokoyama}}, \bibinfo {author} {\bibfnamefont {J.}~\bibnamefont {Linder}}, \
  and\ \bibinfo {author} {\bibfnamefont {A.}~\bibnamefont {Sudb{\o}}},\
  }\href@noop {} {\bibfield  {journal} {\bibinfo  {journal} {Phys. Rev. B}\
  }\textbf {\bibinfo {volume} {77}},\ \bibinfo {pages} {132503} (\bibinfo
  {year} {2008})}\BibitemShut {NoStop}%
\bibitem [{\citenamefont {Beiranvand}\ and\ \citenamefont
  {Hamzehpour}()}]{beiranvand2016spin}%
  \BibitemOpen
  \bibfield  {author} {\bibinfo {author} {\bibfnamefont {R.}~\bibnamefont
  {Beiranvand}}\ and\ \bibinfo {author} {\bibfnamefont {H.}~\bibnamefont
  {Hamzehpour}},\ }\href@noop {} {\bibinfo  {journal} {arXiv:1612.01461}\
  }\BibitemShut {NoStop}%
\bibitem [{\citenamefont {Alomar}\ and\ \citenamefont
  {S{\'a}nchez}(2014)}]{alomar2014thermoelectric}%
  \BibitemOpen
\bibfield  {journal} {  }\bibfield  {author} {\bibinfo {author} {\bibfnamefont
  {M.}~\bibnamefont {Alomar}}\ and\ \bibinfo {author} {\bibfnamefont
  {D.}~\bibnamefont {S{\'a}nchez}},\ }\href@noop {} {\bibfield  {journal}
  {\bibinfo  {journal} {Phys. Rev. B}\ }\textbf {\bibinfo {volume} {89}},\
  \bibinfo {pages} {115422} (\bibinfo {year} {2014})}\BibitemShut {NoStop}%
\bibitem [{\citenamefont {Blundell}\ and\ \citenamefont
  {Blundell}(2009)}]{blundell2009concepts}%
  \BibitemOpen
  \bibfield  {author} {\bibinfo {author} {\bibfnamefont {S.~J.}\ \bibnamefont
  {Blundell}}\ and\ \bibinfo {author} {\bibfnamefont {K.~M.}\ \bibnamefont
  {Blundell}},\ }\href@noop {} {\emph {\bibinfo {title} {Concepts in thermal
  physics}}}\ (\bibinfo  {publisher} {OUP Oxford},\ \bibinfo {year}
  {2009})\BibitemShut {NoStop}%
\bibitem [{\citenamefont {Hwang}\ \emph
  {et~al.}(2016{\natexlab{a}})\citenamefont {Hwang}, \citenamefont {Lopez},\
  and\ \citenamefont {Sanchez}}]{hwang2015large}%
  \BibitemOpen
  \bibfield  {author} {\bibinfo {author} {\bibfnamefont {S.-Y.}\ \bibnamefont
  {Hwang}}, \bibinfo {author} {\bibfnamefont {R.}~\bibnamefont {Lopez}}, \ and\
  \bibinfo {author} {\bibfnamefont {D.}~\bibnamefont {Sanchez}},\ }\href@noop
  {} {\bibfield  {journal} {\bibinfo  {journal} {Phys. Rev. B}\ }\textbf
  {\bibinfo {volume} {94}},\ \bibinfo {pages} {054506} (\bibinfo {year}
  {2016}{\natexlab{a}})}\BibitemShut {NoStop}%
\bibitem [{\citenamefont {Snyder}\ and\ \citenamefont
  {Toberer}(2008)}]{snyder2008complex}%
  \BibitemOpen
  \bibfield  {author} {\bibinfo {author} {\bibfnamefont {G.~J.}\ \bibnamefont
  {Snyder}}\ and\ \bibinfo {author} {\bibfnamefont {E.~S.}\ \bibnamefont
  {Toberer}},\ }\href@noop {} {\bibfield  {journal} {\bibinfo  {journal} {Nat.
  Mater.}\ }\textbf {\bibinfo {volume} {7}},\ \bibinfo {pages} {105} (\bibinfo
  {year} {2008})}\BibitemShut {NoStop}%
\bibitem [{\citenamefont {Sevin{\c{c}}li}\ and\ \citenamefont
  {Cuniberti}(2010)}]{sevinccli2010enhanced}%
  \BibitemOpen
  \bibfield  {author} {\bibinfo {author} {\bibfnamefont {H.}~\bibnamefont
  {Sevin{\c{c}}li}}\ and\ \bibinfo {author} {\bibfnamefont {G.}~\bibnamefont
  {Cuniberti}},\ }\href@noop {} {\bibfield  {journal} {\bibinfo  {journal}
  {Phys. Rev. B}\ }\textbf {\bibinfo {volume} {81}},\ \bibinfo {pages} {113401}
  (\bibinfo {year} {2010})}\BibitemShut {NoStop}%
\bibitem [{\citenamefont {Zebarjadi}\ \emph {et~al.}(2012)\citenamefont
  {Zebarjadi}, \citenamefont {Esfarjani}, \citenamefont {Dresselhaus},
  \citenamefont {Ren},\ and\ \citenamefont {Chen}}]{zebarjadi2012perspectives}%
  \BibitemOpen
  \bibfield  {author} {\bibinfo {author} {\bibfnamefont {M.}~\bibnamefont
  {Zebarjadi}}, \bibinfo {author} {\bibfnamefont {K.}~\bibnamefont
  {Esfarjani}}, \bibinfo {author} {\bibfnamefont {M.}~\bibnamefont
  {Dresselhaus}}, \bibinfo {author} {\bibfnamefont {Z.}~\bibnamefont {Ren}}, \
  and\ \bibinfo {author} {\bibfnamefont {G.}~\bibnamefont {Chen}},\ }\href@noop
  {} {\bibfield  {journal} {\bibinfo  {journal} {Energy \& Environmental
  Science}\ }\textbf {\bibinfo {volume} {5}},\ \bibinfo {pages} {5147}
  (\bibinfo {year} {2012})}\BibitemShut {NoStop}%
\bibitem [{\citenamefont {Giazotto}\ \emph {et~al.}(2006)\citenamefont
  {Giazotto}, \citenamefont {Heikkil{\"a}}, \citenamefont {Luukanen},
  \citenamefont {Savin},\ and\ \citenamefont
  {Pekola}}]{giazotto2006opportunities}%
  \BibitemOpen
  \bibfield  {author} {\bibinfo {author} {\bibfnamefont {F.}~\bibnamefont
  {Giazotto}}, \bibinfo {author} {\bibfnamefont {T.~T.}\ \bibnamefont
  {Heikkil{\"a}}}, \bibinfo {author} {\bibfnamefont {A.}~\bibnamefont
  {Luukanen}}, \bibinfo {author} {\bibfnamefont {A.~M.}\ \bibnamefont {Savin}},
  \ and\ \bibinfo {author} {\bibfnamefont {J.~P.}\ \bibnamefont {Pekola}},\
  }\href@noop {} {\bibfield  {journal} {\bibinfo  {journal} {Rev. Mod. Phys.}\
  }\textbf {\bibinfo {volume} {78}},\ \bibinfo {pages} {217} (\bibinfo {year}
  {2006})}\BibitemShut {NoStop}%
\bibitem [{\citenamefont {Goldsmid}()}]{goldsmid3electronic}%
  \BibitemOpen
  \bibfield  {author} {\bibinfo {author} {\bibfnamefont {H.}~\bibnamefont
  {Goldsmid}},\ }\href@noop {} {\emph {\bibinfo {title} {Electronic
  Refrigeration (Pion, London, 1986)}}},\ Vol.~\bibinfo {volume} {3},\ pp.\
  \bibinfo {pages} {57--87}\BibitemShut {NoStop}%
\bibitem [{\citenamefont {Xu}\ \emph {et~al.}(2014)\citenamefont {Xu},
  \citenamefont {Gan},\ and\ \citenamefont {Zhang}}]{xu2014enhanced}%
  \BibitemOpen
  \bibfield  {author} {\bibinfo {author} {\bibfnamefont {Y.}~\bibnamefont
  {Xu}}, \bibinfo {author} {\bibfnamefont {Z.}~\bibnamefont {Gan}}, \ and\
  \bibinfo {author} {\bibfnamefont {S.-C.}\ \bibnamefont {Zhang}},\ }\href@noop
  {} {\bibfield  {journal} {\bibinfo  {journal} {Phys. Rev. Lett.}\ }\textbf
  {\bibinfo {volume} {112}},\ \bibinfo {pages} {226801} (\bibinfo {year}
  {2014})}\BibitemShut {NoStop}%
\bibitem [{\citenamefont {Liu}\ \emph {et~al.}(2010)\citenamefont {Liu},
  \citenamefont {Sun},\ and\ \citenamefont {Xie}}]{liu2010enhancement}%
  \BibitemOpen
  \bibfield  {author} {\bibinfo {author} {\bibfnamefont {J.}~\bibnamefont
  {Liu}}, \bibinfo {author} {\bibfnamefont {Q.-f.}\ \bibnamefont {Sun}}, \ and\
  \bibinfo {author} {\bibfnamefont {X.}~\bibnamefont {Xie}},\ }\href@noop {}
  {\bibfield  {journal} {\bibinfo  {journal} {Phys. Rev. B}\ }\textbf {\bibinfo
  {volume} {81}},\ \bibinfo {pages} {245323} (\bibinfo {year}
  {2010})}\BibitemShut {NoStop}%
\bibitem [{\citenamefont {Dragoman}\ and\ \citenamefont
  {Dragoman}(2007)}]{dragoman2007giant}%
  \BibitemOpen
  \bibfield  {author} {\bibinfo {author} {\bibfnamefont {D.}~\bibnamefont
  {Dragoman}}\ and\ \bibinfo {author} {\bibfnamefont {M.}~\bibnamefont
  {Dragoman}},\ }\href@noop {} {\bibfield  {journal} {\bibinfo  {journal}
  {Appl. Phys. Lett.}\ }\textbf {\bibinfo {volume} {91}},\ \bibinfo {pages}
  {203116} (\bibinfo {year} {2007})}\BibitemShut {NoStop}%
\bibitem [{\citenamefont {Ohta}\ \emph {et~al.}(2007)\citenamefont {Ohta},
  \citenamefont {Kim}, \citenamefont {Mune}, \citenamefont {Mizoguchi},
  \citenamefont {Nomura}, \citenamefont {Ohta}, \citenamefont {Nomura},
  \citenamefont {Nakanishi}, \citenamefont {Ikuhara}, \citenamefont {Hirano}
  \emph {et~al.}}]{ohta2007giant}%
  \BibitemOpen
  \bibfield  {author} {\bibinfo {author} {\bibfnamefont {H.}~\bibnamefont
  {Ohta}}, \bibinfo {author} {\bibfnamefont {S.}~\bibnamefont {Kim}}, \bibinfo
  {author} {\bibfnamefont {Y.}~\bibnamefont {Mune}}, \bibinfo {author}
  {\bibfnamefont {T.}~\bibnamefont {Mizoguchi}}, \bibinfo {author}
  {\bibfnamefont {K.}~\bibnamefont {Nomura}}, \bibinfo {author} {\bibfnamefont
  {S.}~\bibnamefont {Ohta}}, \bibinfo {author} {\bibfnamefont {T.}~\bibnamefont
  {Nomura}}, \bibinfo {author} {\bibfnamefont {Y.}~\bibnamefont {Nakanishi}},
  \bibinfo {author} {\bibfnamefont {Y.}~\bibnamefont {Ikuhara}}, \bibinfo
  {author} {\bibfnamefont {M.}~\bibnamefont {Hirano}},  \emph {et~al.},\
  }\href@noop {} {\bibfield  {journal} {\bibinfo  {journal} {Nat. Mater.}\
  }\textbf {\bibinfo {volume} {6}},\ \bibinfo {pages} {129} (\bibinfo {year}
  {2007})}\BibitemShut {NoStop}%
\bibitem [{\citenamefont {Linder}\ and\ \citenamefont
  {Bathen}(2016)}]{linder2016}%
  \BibitemOpen
  \bibfield  {author} {\bibinfo {author} {\bibfnamefont {J.}~\bibnamefont
  {Linder}}\ and\ \bibinfo {author} {\bibfnamefont {M.~E.}\ \bibnamefont
  {Bathen}},\ }\href@noop {} {\bibfield  {journal} {\bibinfo  {journal} {Phys.
  Rev. B}\ }\textbf {\bibinfo {volume} {93}},\ \bibinfo {pages} {224509}
  (\bibinfo {year} {2016})}\BibitemShut {NoStop}%
\bibitem [{\citenamefont {Bathen}\ and\ \citenamefont {Linder}()}]{bathen2016}%
  \BibitemOpen
  \bibfield  {author} {\bibinfo {author} {\bibfnamefont {M.~E.}\ \bibnamefont
  {Bathen}}\ and\ \bibinfo {author} {\bibfnamefont {J.}~\bibnamefont
  {Linder}},\ }\href@noop {} {\bibinfo  {journal} {arXiv:1608.06285}\
  }\BibitemShut {NoStop}%
\bibitem [{\citenamefont {Linder}\ and\ \citenamefont
  {Sudb{\o}}(2007)}]{linder2007spin}%
  \BibitemOpen
\bibfield  {journal} {  }\bibfield  {author} {\bibinfo {author} {\bibfnamefont
  {J.}~\bibnamefont {Linder}}\ and\ \bibinfo {author} {\bibfnamefont
  {A.}~\bibnamefont {Sudb{\o}}},\ }\href@noop {} {\bibfield  {journal}
  {\bibinfo  {journal} {Phys. Rev. B}\ }\textbf {\bibinfo {volume} {76}},\
  \bibinfo {pages} {214508} (\bibinfo {year} {2007})}\BibitemShut {NoStop}%
\bibitem [{\citenamefont {Hwang}\ \emph
  {et~al.}(2016{\natexlab{b}})\citenamefont {Hwang}, \citenamefont {Lopez},\
  and\ \citenamefont {Sanchez}}]{hwang}%
  \BibitemOpen
  \bibfield  {author} {\bibinfo {author} {\bibfnamefont {S.-Y.}\ \bibnamefont
  {Hwang}}, \bibinfo {author} {\bibfnamefont {R.}~\bibnamefont {Lopez}}, \ and\
  \bibinfo {author} {\bibfnamefont {D.}~\bibnamefont {Sanchez}},\ }\href@noop
  {} {\bibfield  {journal} {\bibinfo  {journal} {Phys. Rev. B}\ }\textbf
  {\bibinfo {volume} {94}},\ \bibinfo {pages} {054506} (\bibinfo {year}
  {2016}{\natexlab{b}})}\BibitemShut {NoStop}%
\bibitem [{\citenamefont {Alidoust}\ \emph {et~al.}(2010)\citenamefont
  {Alidoust}, \citenamefont {Rashedi}, \citenamefont {Linder},\ and\
  \citenamefont {Sudb{\o}}}]{alidoust2010phase}%
  \BibitemOpen
  \bibfield  {author} {\bibinfo {author} {\bibfnamefont {M.}~\bibnamefont
  {Alidoust}}, \bibinfo {author} {\bibfnamefont {G.}~\bibnamefont {Rashedi}},
  \bibinfo {author} {\bibfnamefont {J.}~\bibnamefont {Linder}}, \ and\ \bibinfo
  {author} {\bibfnamefont {A.}~\bibnamefont {Sudb{\o}}},\ }\href@noop {}
  {\bibfield  {journal} {\bibinfo  {journal} {Phys. Rev. B}\ }\textbf {\bibinfo
  {volume} {82}},\ \bibinfo {pages} {014532} (\bibinfo {year}
  {2010})}\BibitemShut {NoStop}%
\bibitem [{\citenamefont {Zhao}\ \emph {et~al.}(2003)\citenamefont {Zhao},
  \citenamefont {L{\"o}fwander},\ and\ \citenamefont {Sauls}}]{zhao2003phase}%
  \BibitemOpen
  \bibfield  {author} {\bibinfo {author} {\bibfnamefont {E.}~\bibnamefont
  {Zhao}}, \bibinfo {author} {\bibfnamefont {T.}~\bibnamefont {L{\"o}fwander}},
  \ and\ \bibinfo {author} {\bibfnamefont {J.}~\bibnamefont {Sauls}},\
  }\href@noop {} {\bibfield  {journal} {\bibinfo  {journal} {Phys. Rev. Lett.}\
  }\textbf {\bibinfo {volume} {91}},\ \bibinfo {pages} {077003} (\bibinfo
  {year} {2003})}\BibitemShut {NoStop}%
\bibitem [{\citenamefont {Peltonen}\ \emph {et~al.}(2010)\citenamefont
  {Peltonen}, \citenamefont {Virtanen}, \citenamefont {Meschke}, \citenamefont
  {Koski}, \citenamefont {Heikkil{\"a}},\ and\ \citenamefont
  {Pekola}}]{peltonen2010}%
  \BibitemOpen
  \bibfield  {author} {\bibinfo {author} {\bibfnamefont {J.}~\bibnamefont
  {Peltonen}}, \bibinfo {author} {\bibfnamefont {P.}~\bibnamefont {Virtanen}},
  \bibinfo {author} {\bibfnamefont {M.}~\bibnamefont {Meschke}}, \bibinfo
  {author} {\bibfnamefont {J.}~\bibnamefont {Koski}}, \bibinfo {author}
  {\bibfnamefont {T.}~\bibnamefont {Heikkil{\"a}}}, \ and\ \bibinfo {author}
  {\bibfnamefont {J.~P.}\ \bibnamefont {Pekola}},\ }\href@noop {} {\bibfield
  {journal} {\bibinfo  {journal} {Phys. Rev. Lett.}\ }\textbf {\bibinfo
  {volume} {105}},\ \bibinfo {pages} {097004} (\bibinfo {year}
  {2010})}\BibitemShut {NoStop}%
\bibitem [{\citenamefont {Datta}\ and\ \citenamefont
  {Das}(1990)}]{datta1990electronic}%
  \BibitemOpen
  \bibfield  {author} {\bibinfo {author} {\bibfnamefont {S.}~\bibnamefont
  {Datta}}\ and\ \bibinfo {author} {\bibfnamefont {B.}~\bibnamefont {Das}},\
  }\href@noop {} {\bibfield  {journal} {\bibinfo  {journal} {Appl. Phys.
  Lett.}\ }\textbf {\bibinfo {volume} {56}},\ \bibinfo {pages} {665} (\bibinfo
  {year} {1990})}\BibitemShut {NoStop}%
\bibitem [{\citenamefont {H{\"o}gl}\ \emph {et~al.}(2015)\citenamefont
  {H{\"o}gl}, \citenamefont {Matos-Abiague}, \citenamefont {{\v{Z}}uti{\'c}},\
  and\ \citenamefont {Fabian}}]{hogl}%
  \BibitemOpen
  \bibfield  {author} {\bibinfo {author} {\bibfnamefont {P.}~\bibnamefont
  {H{\"o}gl}}, \bibinfo {author} {\bibfnamefont {A.}~\bibnamefont
  {Matos-Abiague}}, \bibinfo {author} {\bibfnamefont {I.}~\bibnamefont
  {{\v{Z}}uti{\'c}}}, \ and\ \bibinfo {author} {\bibfnamefont {J.}~\bibnamefont
  {Fabian}},\ }\href@noop {} {\bibfield  {journal} {\bibinfo  {journal} {Phys.
  Rev. Lett.}\ }\textbf {\bibinfo {volume} {115}},\ \bibinfo {pages} {116601}
  (\bibinfo {year} {2015})}\BibitemShut {NoStop}%
\bibitem [{\citenamefont {Costa}\ \emph {et~al.}(2017)\citenamefont {Costa},
  \citenamefont {H\"{o}gl},\ and\ \citenamefont {Fabian}}]{costa16}%
  \BibitemOpen
  \bibfield  {author} {\bibinfo {author} {\bibfnamefont {A.}~\bibnamefont
  {Costa}}, \bibinfo {author} {\bibfnamefont {P.}~\bibnamefont {H\"{o}gl}}, \
  and\ \bibinfo {author} {\bibfnamefont {J.}~\bibnamefont {Fabian}},\
  }\href@noop {} {\bibfield  {journal} {\bibinfo  {journal} {Phys. Rev. B}\
  }\textbf {\bibinfo {volume} {95}},\ \bibinfo {pages} {024514} (\bibinfo
  {year} {2017})}\BibitemShut {NoStop}%
\bibitem [{\citenamefont {Rashba}(2006)}]{rashba}%
  \BibitemOpen
  \bibfield  {author} {\bibinfo {author} {\bibfnamefont {E.~I.}\ \bibnamefont
  {Rashba}},\ }\href@noop {} {\bibfield  {journal} {\bibinfo  {journal}
  {Physica E}\ }\textbf {\bibinfo {volume} {34}},\ \bibinfo {pages} {31}
  (\bibinfo {year} {2006})}\BibitemShut {NoStop}%
\bibitem [{\citenamefont {Bychkov}\ and\ \citenamefont
  {Rashba}(1984)}]{rashba2}%
  \BibitemOpen
  \bibfield  {author} {\bibinfo {author} {\bibfnamefont {Y.~A.}\ \bibnamefont
  {Bychkov}}\ and\ \bibinfo {author} {\bibfnamefont {E.~I.}\ \bibnamefont
  {Rashba}},\ }\href@noop {} {\bibfield  {journal} {\bibinfo  {journal} {J.
  Phys. C: Solid State Phys.}\ }\textbf {\bibinfo {volume} {17}},\ \bibinfo
  {pages} {6039} (\bibinfo {year} {1984})}\BibitemShut {NoStop}%
\bibitem [{\citenamefont {Sun}\ and\ \citenamefont
  {Shah}(2015)}]{sun2015general}%
  \BibitemOpen
  \bibfield  {author} {\bibinfo {author} {\bibfnamefont {K.}~\bibnamefont
  {Sun}}\ and\ \bibinfo {author} {\bibfnamefont {N.}~\bibnamefont {Shah}},\
  }\href@noop {} {\bibfield  {journal} {\bibinfo  {journal} {Phys. Rev. B}\
  }\textbf {\bibinfo {volume} {91}},\ \bibinfo {pages} {144508} (\bibinfo
  {year} {2015})}\BibitemShut {NoStop}%
\bibitem [{\citenamefont {Zutik}\ \emph {et~al.}(2004)\citenamefont {Zutik},
  \citenamefont {Fabian},\ and\ \citenamefont {Sharma}}]{review}%
  \BibitemOpen
  \bibfield  {author} {\bibinfo {author} {\bibfnamefont {I.}~\bibnamefont
  {Zutik}}, \bibinfo {author} {\bibfnamefont {J.}~\bibnamefont {Fabian}}, \
  and\ \bibinfo {author} {\bibfnamefont {S.~D.}\ \bibnamefont {Sharma}},\
  }\href@noop {} {\bibfield  {journal} {\bibinfo  {journal} {Rev. Mod. Phys.}\
  }\textbf {\bibinfo {volume} {76}},\ \bibinfo {pages} {323} (\bibinfo {year}
  {2004})}\BibitemShut {NoStop}%
\bibitem [{\citenamefont {Blonder}\ \emph {et~al.}(1982)\citenamefont
  {Blonder}, \citenamefont {Tinkham},\ and\ \citenamefont
  {Klapwijk}}]{blonder}%
  \BibitemOpen
  \bibfield  {author} {\bibinfo {author} {\bibfnamefont {G.~E.}\ \bibnamefont
  {Blonder}}, \bibinfo {author} {\bibfnamefont {M.}~\bibnamefont {Tinkham}}, \
  and\ \bibinfo {author} {\bibfnamefont {T.~M.}\ \bibnamefont {Klapwijk}},\
  }\href@noop {} {\bibfield  {journal} {\bibinfo  {journal} {Phys. Rev. B}\
  }\textbf {\bibinfo {volume} {25}},\ \bibinfo {pages} {4515} (\bibinfo {year}
  {1982})}\BibitemShut {NoStop}%
\bibitem [{\citenamefont {{\v{Z}}uti{\'c}}\ and\ \citenamefont
  {Valls}(2000)}]{vzutic2000tunneling}%
  \BibitemOpen
  \bibfield  {author} {\bibinfo {author} {\bibfnamefont {I.}~\bibnamefont
  {{\v{Z}}uti{\'c}}}\ and\ \bibinfo {author} {\bibfnamefont {O.~T.}\
  \bibnamefont {Valls}},\ }\href@noop {} {\bibfield  {journal} {\bibinfo
  {journal} {Phys. Rev. B}\ }\textbf {\bibinfo {volume} {61}},\ \bibinfo
  {pages} {1555} (\bibinfo {year} {2000})}\BibitemShut {NoStop}%
\bibitem [{\citenamefont {De~Gennes}(1999)}]{de1999superconductivity}%
  \BibitemOpen
  \bibfield  {author} {\bibinfo {author} {\bibfnamefont {P.-G.}\ \bibnamefont
  {De~Gennes}},\ }\href@noop {} {\emph {\bibinfo {title} {Superconductivity of
  metals and alloys (advanced book classics)}}}\ (\bibinfo  {publisher}
  {Perseus Books Group},\ \bibinfo {year} {1999})\BibitemShut {NoStop}%
\bibitem [{\citenamefont {Stoner}(1939)}]{stoner1939collective}%
  \BibitemOpen
  \bibfield  {author} {\bibinfo {author} {\bibfnamefont {E.~C.}\ \bibnamefont
  {Stoner}},\ }\href@noop {} {\bibfield  {journal} {\bibinfo  {journal} {Proc.
  R. Soc. A}\ }\textbf {\bibinfo {volume} {169}},\ \bibinfo {pages} {339}
  (\bibinfo {year} {1939})}\BibitemShut {NoStop}%
\bibitem [{\citenamefont {Matos-Abiague}\ \emph {et~al.}(2009)\citenamefont
  {Matos-Abiague}, \citenamefont {Gmitra},\ and\ \citenamefont
  {Fabian}}]{matos2009angular}%
  \BibitemOpen
  \bibfield  {author} {\bibinfo {author} {\bibfnamefont {A.}~\bibnamefont
  {Matos-Abiague}}, \bibinfo {author} {\bibfnamefont {M.}~\bibnamefont
  {Gmitra}}, \ and\ \bibinfo {author} {\bibfnamefont {J.}~\bibnamefont
  {Fabian}},\ }\href@noop {} {\bibfield  {journal} {\bibinfo  {journal} {Phys.
  Rev. B}\ }\textbf {\bibinfo {volume} {80}},\ \bibinfo {pages} {045312}
  (\bibinfo {year} {2009})}\BibitemShut {NoStop}%
\bibitem [{\citenamefont {De~Jong}\ and\ \citenamefont
  {Beenakker}(1995)}]{de1995andreev}%
  \BibitemOpen
  \bibfield  {author} {\bibinfo {author} {\bibfnamefont {M.}~\bibnamefont
  {De~Jong}}\ and\ \bibinfo {author} {\bibfnamefont {C.}~\bibnamefont
  {Beenakker}},\ }\href@noop {} {\bibfield  {journal} {\bibinfo  {journal}
  {Phys. Rev. Lett.}\ }\textbf {\bibinfo {volume} {74}},\ \bibinfo {pages}
  {1657} (\bibinfo {year} {1995})}\BibitemShut {NoStop}%
\bibitem [{\citenamefont {Cao}\ \emph {et~al.}(2004)\citenamefont {Cao},
  \citenamefont {Shi}, \citenamefont {Song}, \citenamefont {Zhou},\ and\
  \citenamefont {Chen}}]{cao2004spin}%
  \BibitemOpen
  \bibfield  {author} {\bibinfo {author} {\bibfnamefont {X.}~\bibnamefont
  {Cao}}, \bibinfo {author} {\bibfnamefont {Y.}~\bibnamefont {Shi}}, \bibinfo
  {author} {\bibfnamefont {X.}~\bibnamefont {Song}}, \bibinfo {author}
  {\bibfnamefont {S.}~\bibnamefont {Zhou}}, \ and\ \bibinfo {author}
  {\bibfnamefont {H.}~\bibnamefont {Chen}},\ }\href@noop {} {\bibfield
  {journal} {\bibinfo  {journal} {Phys. Rev. B}\ }\textbf {\bibinfo {volume}
  {70}},\ \bibinfo {pages} {235341} (\bibinfo {year} {2004})}\BibitemShut
  {NoStop}%
\bibitem [{\citenamefont {Eschrig}(2011)}]{eschrig2011spin}%
  \BibitemOpen
  \bibfield  {author} {\bibinfo {author} {\bibfnamefont {M.}~\bibnamefont
  {Eschrig}},\ }\href@noop {} {\bibfield  {journal} {\bibinfo  {journal} {Phys.
  Today}\ }\textbf {\bibinfo {volume} {64}},\ \bibinfo {pages} {43} (\bibinfo
  {year} {2011})}\BibitemShut {NoStop}%
\bibitem [{\citenamefont
  {Wysoki{\'n}ski}(2012)}]{wysokinski2012thermoelectric}%
  \BibitemOpen
  \bibfield  {author} {\bibinfo {author} {\bibfnamefont {M.}~\bibnamefont
  {Wysoki{\'n}ski}},\ }\href@noop {} {\bibfield  {journal} {\bibinfo  {journal}
  {Acta Phys. Polo. A.}\ }\textbf {\bibinfo {volume} {122}} (\bibinfo {year}
  {2012})}\BibitemShut {NoStop}%
\bibitem [{\citenamefont {Bardas}\ and\ \citenamefont
  {Averin}(1995)}]{bardas1995peltier}%
  \BibitemOpen
  \bibfield  {author} {\bibinfo {author} {\bibfnamefont {A.}~\bibnamefont
  {Bardas}}\ and\ \bibinfo {author} {\bibfnamefont {D.}~\bibnamefont
  {Averin}},\ }\href@noop {} {\bibfield  {journal} {\bibinfo  {journal} {Phys.
  Rev. B}\ }\textbf {\bibinfo {volume} {52}},\ \bibinfo {pages} {12873}
  (\bibinfo {year} {1995})}\BibitemShut {NoStop}%
\bibitem [{\citenamefont {Andreev}(1964{\natexlab{b}})}]{andreev}%
  \BibitemOpen
  \bibfield  {author} {\bibinfo {author} {\bibfnamefont {A.~F.}\ \bibnamefont
  {Andreev}},\ }\href@noop {} {\bibfield  {journal} {\bibinfo  {journal} {Sov.
  Phys. JETP}\ }\textbf {\bibinfo {volume} {19}},\ \bibinfo {pages} {1228}
  (\bibinfo {year} {1964}{\natexlab{b}})}\BibitemShut {NoStop}%
\bibitem [{\citenamefont {Seviour}\ and\ \citenamefont
  {Volkov}(2000)}]{seviour2000giant}%
  \BibitemOpen
  \bibfield  {author} {\bibinfo {author} {\bibfnamefont {R.}~\bibnamefont
  {Seviour}}\ and\ \bibinfo {author} {\bibfnamefont {A.}~\bibnamefont
  {Volkov}},\ }\href@noop {} {\bibfield  {journal} {\bibinfo  {journal} {Phys.
  Rev. B}\ }\textbf {\bibinfo {volume} {62}},\ \bibinfo {pages} {R6116}
  (\bibinfo {year} {2000})}\BibitemShut {NoStop}%
\bibitem [{\citenamefont {Matthias}\ \emph {et~al.}(1961)\citenamefont
  {Matthias}, \citenamefont {Bozorth},\ and\ \citenamefont
  {Van~Vleck}}]{matthias1961ferromagnetic}%
  \BibitemOpen
  \bibfield  {author} {\bibinfo {author} {\bibfnamefont {B.}~\bibnamefont
  {Matthias}}, \bibinfo {author} {\bibfnamefont {R.}~\bibnamefont {Bozorth}}, \
  and\ \bibinfo {author} {\bibfnamefont {J.}~\bibnamefont {Van~Vleck}},\
  }\href@noop {} {\bibfield  {journal} {\bibinfo  {journal} {Phys. Rev. Lett.}\
  }\textbf {\bibinfo {volume} {7}},\ \bibinfo {pages} {160} (\bibinfo {year}
  {1961})}\BibitemShut {NoStop}%
\bibitem [{\citenamefont {Tokuyasu}\ \emph {et~al.}(1988)\citenamefont
  {Tokuyasu}, \citenamefont {Sauls},\ and\ \citenamefont
  {Rainer}}]{tokuyasu1988proximity}%
  \BibitemOpen
  \bibfield  {author} {\bibinfo {author} {\bibfnamefont {T.}~\bibnamefont
  {Tokuyasu}}, \bibinfo {author} {\bibfnamefont {J.}~\bibnamefont {Sauls}}, \
  and\ \bibinfo {author} {\bibfnamefont {D.}~\bibnamefont {Rainer}},\
  }\href@noop {} {\bibfield  {journal} {\bibinfo  {journal} {Phys. Rev. B}\
  }\textbf {\bibinfo {volume} {38}},\ \bibinfo {pages} {8823} (\bibinfo {year}
  {1988})}\BibitemShut {NoStop}%
\bibitem [{\citenamefont {Deng}\ \emph {et~al.}(2012)\citenamefont {Deng},
  \citenamefont {Yu}, \citenamefont {Huang}, \citenamefont {Larsson},
  \citenamefont {Caroff},\ and\ \citenamefont {Xu}}]{deng2012anomalous}%
  \BibitemOpen
  \bibfield  {author} {\bibinfo {author} {\bibfnamefont {M.}~\bibnamefont
  {Deng}}, \bibinfo {author} {\bibfnamefont {C.}~\bibnamefont {Yu}}, \bibinfo
  {author} {\bibfnamefont {G.}~\bibnamefont {Huang}}, \bibinfo {author}
  {\bibfnamefont {M.}~\bibnamefont {Larsson}}, \bibinfo {author} {\bibfnamefont
  {P.}~\bibnamefont {Caroff}}, \ and\ \bibinfo {author} {\bibfnamefont
  {H.}~\bibnamefont {Xu}},\ }\href@noop {} {\bibfield  {journal} {\bibinfo
  {journal} {Nano Lett.}\ }\textbf {\bibinfo {volume} {12}},\ \bibinfo {pages}
  {6414} (\bibinfo {year} {2012})}\BibitemShut {NoStop}%
\bibitem [{\citenamefont {Steeneken}\ \emph {et~al.}(2002)\citenamefont
  {Steeneken}, \citenamefont {Tjeng}, \citenamefont {Elfimov}, \citenamefont
  {Sawatzky}, \citenamefont {Ghiringhelli}, \citenamefont {Brookes},\ and\
  \citenamefont {Huang}}]{Steeneken2002}%
  \BibitemOpen
  \bibfield  {author} {\bibinfo {author} {\bibfnamefont {P.~G.}\ \bibnamefont
  {Steeneken}}, \bibinfo {author} {\bibfnamefont {L.~H.}\ \bibnamefont
  {Tjeng}}, \bibinfo {author} {\bibfnamefont {I.}~\bibnamefont {Elfimov}},
  \bibinfo {author} {\bibfnamefont {G.~A.}\ \bibnamefont {Sawatzky}}, \bibinfo
  {author} {\bibfnamefont {G.}~\bibnamefont {Ghiringhelli}}, \bibinfo {author}
  {\bibfnamefont {N.~B.}\ \bibnamefont {Brookes}}, \ and\ \bibinfo {author}
  {\bibfnamefont {D.~J.}\ \bibnamefont {Huang}},\ }\href@noop {} {\bibfield
  {journal} {\bibinfo  {journal} {Phys. Rev. Lett.}\ }\textbf {\bibinfo
  {volume} {88}},\ \bibinfo {pages} {047201} (\bibinfo {year}
  {2002})}\BibitemShut {NoStop}%
\bibitem [{\citenamefont {Santos}\ and\ \citenamefont
  {Moodera}(2004)}]{Tiffany2004}%
  \BibitemOpen
  \bibfield  {author} {\bibinfo {author} {\bibfnamefont {T.~S.}\ \bibnamefont
  {Santos}}\ and\ \bibinfo {author} {\bibfnamefont {J.~S.}\ \bibnamefont
  {Moodera}},\ }\href@noop {} {\bibfield  {journal} {\bibinfo  {journal} {Phys.
  Rev. B}\ }\textbf {\bibinfo {volume} {69}},\ \bibinfo {pages} {241203(R)}
  (\bibinfo {year} {2004})}\BibitemShut {NoStop}%
\bibitem [{\citenamefont {Manchon}\ \emph {et~al.}(2015)\citenamefont
  {Manchon}, \citenamefont {Koo}, \citenamefont {Nitta}, \citenamefont
  {Frolov},\ and\ \citenamefont {Duine}}]{manchon3}%
  \BibitemOpen
  \bibfield  {author} {\bibinfo {author} {\bibfnamefont {A.}~\bibnamefont
  {Manchon}}, \bibinfo {author} {\bibfnamefont {H.~C.}\ \bibnamefont {Koo}},
  \bibinfo {author} {\bibfnamefont {J.}~\bibnamefont {Nitta}}, \bibinfo
  {author} {\bibfnamefont {S.~M.}\ \bibnamefont {Frolov}}, \ and\ \bibinfo
  {author} {\bibfnamefont {R.~A.}\ \bibnamefont {Duine}},\ }\href@noop {}
  {\bibfield  {journal} {\bibinfo  {journal} {Nat. Mater.}\ }\textbf {\bibinfo
  {volume} {14}},\ \bibinfo {pages} {871} (\bibinfo {year} {2015})}\BibitemShut
  {NoStop}%
\bibitem [{\citenamefont {Huang}\ \emph {et~al.}(2011)\citenamefont {Huang},
  \citenamefont {Badrutdinov}, \citenamefont {Serra}, \citenamefont {Kodera},
  \citenamefont {Nakaoka}, \citenamefont {Kumagai}, \citenamefont {Arakawa},
  \citenamefont {Tayurskii}, \citenamefont {Kono},\ and\ \citenamefont
  {Ono}}]{Huang2011}%
  \BibitemOpen
  \bibfield  {author} {\bibinfo {author} {\bibfnamefont {S.~M.}\ \bibnamefont
  {Huang}}, \bibinfo {author} {\bibfnamefont {A.~O.}\ \bibnamefont
  {Badrutdinov}}, \bibinfo {author} {\bibfnamefont {L.}~\bibnamefont {Serra}},
  \bibinfo {author} {\bibfnamefont {T.}~\bibnamefont {Kodera}}, \bibinfo
  {author} {\bibfnamefont {T.}~\bibnamefont {Nakaoka}}, \bibinfo {author}
  {\bibfnamefont {N.}~\bibnamefont {Kumagai}}, \bibinfo {author} {\bibfnamefont
  {Y.}~\bibnamefont {Arakawa}}, \bibinfo {author} {\bibfnamefont {D.~A.}\
  \bibnamefont {Tayurskii}}, \bibinfo {author} {\bibfnamefont {K.}~\bibnamefont
  {Kono}}, \ and\ \bibinfo {author} {\bibfnamefont {K.}~\bibnamefont {Ono}},\
  }\href@noop {} {\bibfield  {journal} {\bibinfo  {journal} {Phys. Rev. B}\
  }\textbf {\bibinfo {volume} {84}},\ \bibinfo {pages} {085325} (\bibinfo
  {year} {2011})}\BibitemShut {NoStop}%
\bibitem [{\citenamefont
  {Onsager}(1931{\natexlab{a}})}]{onsager1931reciprocal1}%
  \BibitemOpen
  \bibfield  {author} {\bibinfo {author} {\bibfnamefont {L.}~\bibnamefont
  {Onsager}},\ }\href@noop {} {\bibfield  {journal} {\bibinfo  {journal} {Phys.
  Rev.}\ }\textbf {\bibinfo {volume} {37}},\ \bibinfo {pages} {405} (\bibinfo
  {year} {1931}{\natexlab{a}})}\BibitemShut {NoStop}%
\bibitem [{\citenamefont
  {Onsager}(1931{\natexlab{b}})}]{onsager1931reciprocal2}%
  \BibitemOpen
  \bibfield  {author} {\bibinfo {author} {\bibfnamefont {L.}~\bibnamefont
  {Onsager}},\ }\href@noop {} {\bibfield  {journal} {\bibinfo  {journal} {Phys.
  Rev.}\ }\textbf {\bibinfo {volume} {38}},\ \bibinfo {pages} {2265} (\bibinfo
  {year} {1931}{\natexlab{b}})}\BibitemShut {NoStop}%
\end{thebibliography}%
\end{document}